\documentclass[pre,onecolumn]{revtex4}
\usepackage{graphicx}
\usepackage{amsopn}
\usepackage{amsfonts}
\usepackage{latexsym}
\usepackage{amsmath,bm,amssymb}

\usepackage{color}
\usepackage{graphicx}
\usepackage[T2A]{fontenc}

\begin{document}

\title{Dynamic flexoelectric instabilities in nematic liquid crystals}

\author{E.S. Pikina$^{1,2}$, A.R. Muratov$^2$, E.I. Kats$^1$, and V.V. Lebedev$^{1,3}$.}

\affiliation{$^1$ Landau Institute for Theoretical Physics, RAS, \\
142432, Chernogolovka, Moscow region, Russia, \\
$^2$ Institute for Oil and Gas Research, RAS, 119917, Gubkina 3, Moscow, Russia,\\
$^3$ NRU Higher School of Economics, \\
101000, Myasnitskaya 20, Moscow, Russia.}

\begin{abstract}

Electro-hydrodynamic phenomena in liquid crystals constitute an old but still very active research area. The reason is that these phenomena play the key role in various applications of liquid crystals and due to the general interest of physical community to out-of-equilibrium systems. Nematic liquid crystals (NLCs) are ideally representative for such investigations. Our article aims to study theoretically the linear NLCs dynamics. We include into consideration orientation elastic energy, hydrodynamic motion, external alternating electric field, electric conductivity and flexoelectric polarization. We analyze the linear stability of the NLC film, determining dynamics of perturbations with respect to the homogeneous initial state of the NLC. For the purpose we compute eigen-values of the evolution matrix for a period of the external alternating electric field. These eigen-values determine the amplification factors for the modes during the period. The instability occurs when the principal eigen-value of the evolution matrix becomes unity by its absolute value. The condition determines the threshold (critical field) for the instability of the uniform state. It turns out  that one might expect various types of the instability, only partially known and investigated in the literature. Particularly, we find that the flexoelectric instability may lead to two-dimensionally space modulated patterns exhibiting time oscillations. This type of the structures was somehow overlooked in the previous works. We formulate conditions needed for the scenario to be realized. We hope that the results of our work will open the door to a broad range of further studies. Of especial importance would be a comprehensive understanding of the role of various material parameters and non-linear effects which is a key step for the rational design of NLCs exhibiting the predicted in this publication multidimensional oscillating in time patterns.

\end{abstract}


\maketitle

\section{Introduction}
\label{sec:intro}

A wide variety of pattern-forming instabilities in NLCs under the influence of electric field has been extensively investigated already about 50 years, see, e.g., Refs.
\cite{WI63,HZ68,HE69,ML77,RS88,GP89,RS89,PI91,Meyer69,Pikin,Barnik1978,Pikin1979,Pikin2018,RB91,BR91,ZI91,CH92,GP93,BC94,BK96,KH96,KL03,OP05,ZA06,CG09,TK14,KD15}. And many more references can be added. Therefore, one could think that fundamental studies of this phenomenon are exhausted. However, recently the topic was resurrected and attracted much attention of researchers. At least partially because the field has been enriched by observations of localized and propagating excitations in NLCs under external alternating electric field \cite{lavrentovich1,lavrentovich2,lavrentovich3}. Note that similar phenomena (i.e., localized and propagating excitations) have been observed in different kinds of liquid crystals and under different conditions \cite{dierking1,dierking2,dierking3,dierking4,dierking5,aya}.

The observations of Refs. \cite{lavrentovich1,lavrentovich2,lavrentovich3} were made at the conditions, where the flexoelectric effect plays a crucial role. The flexoelectricity in liquid crystals has been introduced long ago by Meyer and then studied in many works \cite{WI63,HZ68,HE69,ML77,RS88,GP89,RS89,PI91,Meyer69,Pikin,Barnik1978,Pikin1979,Pikin2018,RB91,BR91,ZI91,CH92,GP93,BC94,BK96,KH96,KL03,OP05,ZA06,CG09,TK14,KD15}.
However, the results reported in Refs. \cite{lavrentovich1,lavrentovich2,lavrentovich3,dierking1,dierking2,dierking3,dierking4,dierking5,aya}
suggest that the flexoelectric mechanism can bring about a variety of unknown scenarios. The observed in these works localized and propagating excitations suggest that the mechanism behind is based on non-linear physics in out-of-equilibrium dissipative systems. The first mandatory step to rationalize the observed in these works non-trivial dynamic behavior is to solve the linear dynamic equations for the NLCs in the external a.c. field. This is the aim of our publication. For the problem under consideration even the analysis of solutions of the linearized equations turns out rather tricky. That is why in this paper we have deliberately focused on the most limited questions of linear stability, and postpone the non-linear step for the further works.

In this work we solve numerically the set of the linear electro-hydrodynamic equations in an external alternating electric field. Our scheme includes all essential ingredients of the problem. Namely: the Frank elasticity, the hydrodynamic motion, the electric conductivity and the flexoelectric polarization. Besides we assume the uniform boundary conditions with strong surface anchoring for the director. The hydrodynamic motion is assumed to be incompressible, that the mass density $\rho$ can be regarded to be constant and the incompressibility condition $\nabla \bm v=0$ is imposed on the velocity field $\bm v$. The incompressibility condition is explained by small values of Mach number in the hydrodynamic motion that we are interested in.

The linear electro-hydrodynamic equations for NLCs are known (see, e.g., \cite{BK96,KD15,KN04,KP08,TE08,PK18,EC19,KL93,CL00}). However, in the most of the cited above publications, not all essential ingredients were included into consideration: e.g., hydrodynamic motion or finite (although small) electric conductivity. To be sure that nothing is missing in the previous works (where the linear dynamic equations were constructed by the symmetry arguments) we re-derived the equations by the linearization of the obtained recently \cite{PM23} by our group complete set of non-linear dynamic equations of NLCs. It turns out that our set of the linear equations coincides (up to notations) with the equations presented in Ref. \cite{KP08}. Nevertheless we present our derivation of the equations in the appendix \ref{sec:LinEqs}, and solve the equations in the main body of the paper. Partially because in our approach it is essential to separate explicitly  the contributions from the ideally insulating NLC. Then the finite conductivity effects are included as a perturbation over small conductivity of the NLC.

In view of time periodicity of the external electric field, the equations of nematodynamics have to be solved in terms of Floquet normal form in time $t$. Thus the solution is characterized by a set of characteristic exponents (decrements/increments) $\lambda_i$, see Refs. \cite{Floquet1883,Erugin1966}. We are interested in the dynamic behavior of the NLC on time scales much larger than the period $T$ of the external alternating electric field. The behavior of the amplitudes of the modes on the time scales is described by the factors $\exp(\lambda_i t)$. To establish the set of $\lambda_i$, we solve the equations for one period and find the amplification factors $\exp(\lambda_i T)$ of the eigen-modes of the evolution operator.

The flexoelectric effect makes the state of the nematic with the homogeneous director field unstable in high enough external electric fields. The critical electric field, corresponding to the instability threshold, is determined by the condition of zero real part of the main characteristic exponent, related to the critical mode. All other characteristic exponents have negative real parts at the threshold. Our aim is to perform the linear stability analysis and to examine the behavior of the nematic near the instability threshold. Unfortunately, the equations cannot be solved analytically. That is why we investigate their solutions numerically and supplement the numerics by analytic expressions which can be obtained for the pulse-like time dependent electric field.

We study mainly the simplified model, assuming that all three Frank modules are equal in the director elastic energy, and the only flexo-coefficient is taken into account. Besides we assume that the viscous dissipation energy is characterized by a single viscosity coefficient $\eta$ (see details in the next section). This approach allows us to bury our ignorance about the actual magnitudes of the many material characteristics of nematics in a few phenomenological parameters. However, even the simplified model enables one to draw general conclusions applying to any NLC. We believe that the model reflects correctly physics of the dynamic flexoelectric instability. Unfortunately even the simplified model of the linear electro-hydrodynamics of NLCs contains a large number (often poorly known experimentally) of material parameters. Therefore a complete analysis of such a multiparametric dynamic phase diagram is impossible even numerically (and, most importantly, is not very meaningful). We aim to draw general qualitative conclusions concerning the character of the dynamic flexoelectric instability. To confirm that our qualitative conclusions are robust with respect to the model assumptions, we discuss also some results obtained by going beyond the simplified model.

We encounter competing bifurcations of the initially uniform director field, leading to the following patterns:
(i) stationary stripe structures; (ii) stationary two dimensionally modulated structures; (iii) oscillating in time two dimensional structures. Which one of these bifurcations appears the first upon increasing the external electric field depends on the material parameters and the field frequency $\omega$. Stripe flexoelectric domains, the case (i), are well known \cite{Pikin,Barnik1978,Pikin1979,Pikin2018}. Stationary two dimensional patterns, the case (ii), were predicted and observed experimentally, see Refs. \cite{BK96,KN04,KP08,TE08,PK18}. However, scanning the literature, we did not find theoretical predictions or experimental observations of the oscillating in time states in NLCs.

At the first step we solve the complete set of the linear equations for the unbounded NLC. The differential equations for this case can be solved using the Fourier transform. Then we find the dynamic behavior of the system during a period of the external electric field for different values of the wave vector. As a result of this, relatively simple, step we find the characteristic exponents $\lambda_i$, as functions of the three-component wave vector $\bm q$. At the second step we consider the nematic film restricted by plane plates and use suitable boundary conditions on the surfaces of the plates. To check generality of our qualitative conclusions, we solved the dynamic equations for two simple types of the time-dependent external field. Namely for the harmonic and for the pulsing external electric field (that is the field which is constant during the half of period and changes its sign at the second half of the period). We find that all qualitative features of the solutions are the same for the cases.

We compared the solutions of the equations for the unbounded nematic and a much more computer time consuming solutions of the equations for the finite thickness film. In this case we face with the set of the partial differential equations which cannot be reduced to ODE. Besides, the boundary conditions should be added to the set of the equations, that is not completely trivial. The problem is that for an incompressible fluid its pressure is determined by the incompressibility condition, that is nonlocal. The comparison of the results obtained for the unbounded and for the film of finite thickness, allows us to claim that the results for the cases are in well qualitative agreement (and even in quite satisfactory quantitative agreement). This finding enables one to scan fast a wide interval of the parameters to find the most interesting regions. Then it can be examined in more detail for the film. The peculiarity is based upon the reasoning, that properties of the nematic films of thickness larger than the inverse instability wave vector differ weakly from the unbounded nematic.

Our paper is divided into five sections of which this introduction is the first. In Section \ref{sec:general} we introduce the main relations underlying our analysis and the computational scheme. In Section \ref{sec:P1}, the linear stability of an unbounded nematic in an external alternating electric field is analyzed. We compare the results for the harmonic and for the pulse dependence of the field on time. Section \ref{sec:P2} is devoted to linear stability analysis for a nematic film of finite thickness. The nematodynamic equations are solved with suitable boundary conditions. Our results are outlined in Conclusion. Particularly, we discuss a possibility of the appearance of oscillating in time states above the instability threshold. Appendix \ref{sec:LinEqs} contains a derivation of the linearized equations from the general system of non-linear equations of the nematic electro-hydrodynamics derived in Ref. \cite{PM23}. All technical details needed to analyze the dynamics of flexoelectric instability for finite size system are presented in Appendix \ref{app:finite}. We relegated into Appendix \ref{app:comparison} the tables of the results of our computations to compare quantitatively the results obtained for the unbounded systems (with $q_z \simeq 1/d$) and those for the corresponding finite size $d$ films.

\section{General relations}
\label{sec:general}

Nematic liquid crystals are anisotropic fluids, the anisotropy is described in terms of the director field $\bm n$, that is the unit headless vector. Distortions of the director fields are associated with the nematic elastic energy
\begin{eqnarray}
\int dV \left\{\frac{K_1}{2}(\nabla \bm n)^2
+\frac{K_2}{2}[ \bm n (\nabla\times \bm n)]^2
+ \frac{K_3}{2}[ \bm n \times (\nabla\times \bm n)]^2 \right\},
\label{Franken}
\end{eqnarray}
that is called Frank energy. Here $K_1,K_2,K_3$ are Frank modules, typically of the same order.

The electric energy of the nematics is anisotropic as well. Following the general ideology \cite{LL84,GP93}, it can written as
\begin{equation}
-\frac{\epsilon_0}{2}\int dV\,
\left\{ \epsilon_\parallel (\bm n \bm E)^2
+\epsilon_\perp [E^2-(\bm n \bm E)^2] \right\},
\label{electricen}
\end{equation}
where $\bm E$ is electric field, $\epsilon_\parallel$ and $\epsilon_\perp$ are components of the dielectric permittivity, longitudinal and perpendicular to the director $\bm n$. The degree of the anisotropy is characterized by the difference $\Delta\epsilon=\epsilon_\parallel-\epsilon_\perp$.

In addition, the nematics possess the flexoelectric polarization $\bm P_{fl}$ related to distortions of the director field
\begin{eqnarray}
\bm P_{fl}= e_1 \bm n (\nabla \bm n) + e_3 (\bm n \nabla )\bm n .
\label{weakc7}
\end{eqnarray}
Here $e_1, e_3$ are flexo-coefficients related to splay-like and bend-like deformations of the director field $\bm n$. The flexoelectric energy is of the nematic written as
\begin{equation}
-\int dV\, \bm P_{fl} \bm E.
\label{flexenergy}
\end{equation}
Note that for the homogeneous in space electric field $\bm E$ the flexoelectric contribution (\ref{flexenergy}) into the bulk energy of the nematic (i.e., with surface terms neglected) is determined by the single combination $\zeta=e_1-e_3$.

Dynamics of the nematics is determined by a set of kinetic coefficients. First of all, we should note the director rotational viscosity $\gamma$ determining kinetics of the director. The quantity has the dimension of the hydrodynamic viscosity coefficients and is usually of the same order. Due to the anisotropy there are five independent viscosity coefficients in the nematic, entering the forth order viscosity tensor $\eta_{ijkl}$ \cite{GP93}. The viscous dissipative stress tensor reads as
\begin{eqnarray}
&& \eta_{ijkl} A_{kl} = 2\eta _2 A_{ij}
+ \, 2(\eta_3 - \eta_2)(A_{ik}n_kn_j + A_{jk}n_i n_k)
\nonumber \\
\label{viscosity1}
&& +(\eta_4 - \eta _2)\delta_{ij}A_{kk} +2(\eta_1 + \eta_2 - 2\eta_3) n_in_jn_kn_lA_{kl}
\nonumber
 \\ &&  +
(\eta_5 - \eta_4 + \eta _2)(\delta _{ij}n_k n_lA_{kl} + n_i n_j A_{kk}) \, ,
\nonumber
\end{eqnarray}
where $A_{ij} =\partial _i v_j + \partial _j v_i$. The coefficients are introduced in accordance with Ref. \cite{GP93}. The viscosity coefficients are usually of the same order.

We examine the nematics in the external alternating electric field $E$. First, we consider the harmonically varying in time electric field $E=E_0\cos(\omega t)$ where $E_0$ is the amplitude of the electric field and $\omega$ is its frequency. The field is periodic with the period $T=2\pi/\omega$. Second, we consider the pulsing periodic field consisting of both positive and negative pulses. During a period of duration $T$ its time dependence is
\begin{equation}
E(t)=E_0, \ 0<t<T/2, \quad E(t)=-E_0, \ T/2<t<T\ ,
\label{meand}
\end{equation}
where $E_0$ is the amplitude of the field. The temporal dependence (\ref{meand}) of the external electric field was first considered (within the model neglecting hydrodynamic motion and conductivity) in Ref. \cite{Pikin}.

Nematics are weak electrolytes where both, positive and negative ions, carry the electric current. The density of the electric current $\bm j$ is determined by the conductivity, related to the ions. Due to the anisotropy of nematics the relation between the electric field $\bm E$ and the density of the electric current is
\begin{equation}
\bm j = \sigma_\parallel \bm n (\bm n \bm E)
+\sigma_\perp [\bm E-\bm n (\bm n \bm E)],
\label{conductivity}
\end{equation}
where $\sigma_\parallel$ and $\sigma_\perp$ are the components of the conductivity longitudinal and perpendicular to the director $\bm n$. The conductivity can be estimated as
\begin{equation}
\sigma\sim \frac{c e^2}{k_B T}D,
\nonumber
\end{equation}
where $D$ is the diffusion coefficient of the ions, $c$ is their density, $e$ is the electron charge and $T$ is temperature.

Using the expression for the density of the electric current (\ref{conductivity}), we ignore contributions to $\bm j$ related to possible inhomogeneities of the space distribution of the ions. Therefore the frequency $\omega$ should be much larger than the ions relaxation rate $D/r_D^2$ where Debye length $r_D$ is estimated as
\begin{equation}
r_D\sim \left(\frac{k_B T \epsilon_0 \epsilon_\perp}{ c e^2}\right)^{1/2}.
\nonumber
\end{equation}
Therefore the condition $\omega \gg D/r_D^2$ can be rewritten as
\begin{equation}
\omega \gg \frac{\sigma}{\epsilon_0 \epsilon_\perp}.
\label{frequencycond}
\end{equation}
The inequality (\ref{frequencycond}) is assumed to be satisfied in what follows.

We are interested in the flexoelectric instability. To restrict to only this case, it is convenient to avoid other types of electro-hydrodynamic instability. In this respect one should distinguish materials with positive and negative signs of the dielectric permittivity difference $\Delta\epsilon=\epsilon_\parallel-\epsilon_\perp$ and of the conductivity difference $\Delta\sigma=\sigma_\parallel-\sigma_\perp$. Correspondingly, there are four classes of the NLCs. These classes are designated as $(- -)$, $(- +)$, $(+ -)$, and $(+ +)$, where the first sign in the brackets stands for the dielectric permittivity and the second one stands for the conductivity.

At increasing the external electric field the $(+ +)$ and $(+ -)$ materials experience well-known Frederiksz instability, not related to flexoelectricity. The theoretical analysis of electro-hydrodynamic instabilities in nematic films and the experimental data \cite{RS88,GP89,RS89,PI91,RB91,BR91,ZI91,CH92,GP93,BC94,BK96,CG09,TK14,KD15,KN04,KP08,TE08,PK18} shows that for the $(- +)$ nematics the instability usually leads to static stripes. At some conditions the instability leads to a two-dimensional pattern of the director field \cite{KP08}. However, the pattern appears to be stationary as well. It is why in this paper we consider solely the $(- -)$ materials.

There is a hierarchy of the relaxation times in NLCs. The slowest mode is related to the director relaxation. Its decrement is determined by the director rotational viscosity $\gamma$ and the Frank modules $K_1,K_2,K_3$. The decrement is estimated as $K q^2/\gamma$ where $K$ stands for $K_1,K_2,K_3$ and $q$ is the characteristic wave vector of the considered mode.

The relaxation rate of the hydrodynamic motion of the nematic at a given wave vector $q$ can be estimated as $\eta q^2/\rho$ where $\rho$ is the mass density, and $\eta$ estimates the dynamic viscosity coefficients of the nematic. This relaxation rate is much larger, than the director distortion relaxation rate. The ratio of these rates is the dimensionless parameter $K\rho \gamma^{-1} \eta^{-1}$, independent of $q$. Usually $\gamma\sim \eta$, therefore the parameter can be written as $K \rho \eta^{-2}$. Typically in NLCs it is a small parameter in the range $10^{-4} \div 10^{-3}$ \cite{KD15,JE80,KN04,KP08,TE08,PK18,EC19}. The smallness of the parameter $K \rho \eta^{-2}$ means that at studying dynamics of the director distortions the velocity $\bm v$ of the nematic can be treated in the adiabatic approximation. By other words, the velocity adjusts simultaneously to the director field. The approximation leads to the estimate $v\sim (K q/\eta) \delta n$ for the velocity, where $\delta n$ is the director field distortion.

Comparing Frank energy (\ref{Franken}) and the anisotropy of the electric energy (\ref{electricen}), we find the relation
\begin{equation}
|\Delta\epsilon| \epsilon_0 E_c^2 \sim K q_c^2,
\label{escrit1}
\end{equation}
determining the wave vector $q_c$ of the critical mode. Here $E_c$ is the amplitude of the external electric field corresponding to the onset of the flexoelectric instability. Comparing then the director relaxation rate $\sim K q^2/\gamma$ and the influence of the external electric field on the director dynamics related to the anisotropy of the dielectric permittivity, we arrive at the estimate
\begin{equation}
\vert\Delta\epsilon\vert \epsilon_0 E_c^2\, \sim \, \gamma\,\omega\, .
\label{Ethr}
\end{equation}
Combining the estimates (\ref{escrit1},\ref{Ethr}), we find that
\begin{equation}
q_c \sim (\gamma \omega /K)^{1/2},
\label{escrit2}
\end{equation}
for the critical wave vector.

The flexoelectric energy (\ref{flexenergy}) compete with Frank energy (\ref{Franken}). Comparing the energies for the critical values (\ref{Ethr},\ref{escrit2}) we find that their ratio is determined by the factor $\zeta(|\Delta \epsilon| K)^{1/2}$. Thus no flexoelectric instability occurs if $\zeta$ is too small. The instability, observed at increasing the external alternating electric field, occurs if the inequality is correct
\begin{equation}
\zeta^2/(|\Delta\epsilon|\epsilon_0 K_1) \, >\, C,
 \label{flexcrit}
 \end{equation}
where $C$ is a constant of order unity. Its value is not universal being dependent on the material parameters of the nematic. The condition (\ref{flexcrit}) is a generalization of the analogous condition for the onset of the flexoelectric instability in the static electric field (the same holds for the pulse field within the model neglecting hydrodynamic motion and conductivity, see Refs. \cite{Pikin,Barnik1978,Pikin1979,Pikin2018}).

For the nematic film of finite thickness $d$, the estimates, formulated above, are correct provided $q_c d \gtrsim 1$. Further, the condition is assumed to be satisfied. Moreover, we consider relatively thick films where $q_c d$ is large. The main qualitative findings of our work are valid for such "thick" films, and all quantitative results for the film of finite thickness are close to those for the unbounded sample.

In our numerical computations we use mainly the simplified model with a single Frank module  $K_1=K_2=K_3=K$ and with a single viscosity coefficient $\eta$. Namely, we assume that $\eta_1 = \eta_2 =\eta_3 = \eta_4 \equiv \eta $, and $\eta_5 = 0$, see Eq. (\ref{viscosity1}). It is easy to check that such a choice does not violate the conditions for the positive entropy production, see the conditions in Ref. \cite{GP93}. We assume $e_3=0$ as well. To check generality of the results obtained in the framework of the simplified model we performed also the computations giving up some restrictions of the simplified model, namely, for different Frank modules and for non-zero $e_3$.

In the framework of the simplified model we deal with the following dimensionless parameters
\begin{equation}
\eta/\gamma,  \quad  \Delta\epsilon/\epsilon_\perp, \quad K_1 |\Delta\epsilon| \epsilon_0/\zeta^2,
\label{dparameters}
\end{equation}
controlling the character of the instability. All the subsequent general conclusions can be formulated in terms of the dimensionless parameters.

Note that not all relevant for the NLC dynamic parameters are reliably known experimentally even for the so-to-speak standard (e.g., MBBA) nematics. All the more it is true for some new recently synthesized materials, see, e.g., Refs. \cite{lavrentovich1,lavrentovich2,lavrentovich3,dierking1,dierking2,dierking3,dierking4,dierking5,aya}.
Thus one may play with the material parameters (of course within the limits of physically acceptable values). Following this way, we predict the region of parameters where the new scenario of the instability is realized, related to oscillating in time patterns.

\subsection{Linear dynamic equations of NLCs}
\label{subsec:lineareq}

To be specific, in what follows we consider the nematic film enclosed by two parallel plates and chose the $X$ and $Y$ axes of the reference system along the plates. We assume, that the external alternating electric field is directed along the $Z$ axis, and that without the external electric field director $\bm n$ is aligned everywhere along the $X$ axis. Experimentally, the geometry can be achieved by special preparing the surfaces of the plates guaranteeing that $\bm n$ is directed along $X$ axis there. Above the instability threshold the director field $\bm n$ loses its homogeneity.

Since the nematic film is assumed to be homogeneous in the $X - Y$ plane, one can examine modes with the harmonic dependence $\exp(iq_x x+i q_y y)$ of all varying quantities. The flexoelectric instability occurs at a finite wave vector $\bm q=(q_x,q_y)$. The stripe structure of the director field above the instability threshold corresponds to $q_x=0$ or to $q_y=0$ for the critical mode. If both components of the wave vector of the critical mode are non-zero, then two-dimensional modulated structures can be realized.

We perform the linear stability analysis of the nematic liquid crystal dynamics in the presence of an alternating electric field. For the purpose we use the system of linear equations of the nematic, in the form
derived by linearization of the complete system of non-linear equations \cite{PM23}. Readers interested in technical details can find those in Appendix \ref{sec:LinEqs}.

Within our simplified model ($K_1=K_2=K_3=K$, $e_3=0$ and a single viscous coefficient $\eta$), one finds the following system of equations for the fields with the dependence $\propto \exp(iq_x x+i q_y y)$
\begin{eqnarray}
\partial_t n_y=iq_x v_y +\frac 1{\gamma}
\left[ K(\partial_z^2-q^2) n_y -i\zeta q_y n_z E(t)
-\zeta q_x q_y\Phi \right] \, ,
\label{linear1} \\
\partial_t n_z =iq_x v_z +\frac 1{\gamma}
\left[ K(\partial_z^2-q^2) n_z +i\zeta q_y n_y E(t)
+\Delta {\epsilon }\epsilon_0 E(t)^2 n_z \right.
\nonumber \\
\left. -i\Delta {\epsilon }\epsilon_0 E(t) q_x \Phi
+i\zeta q_x\partial_z \Phi \right] \, ,
\label{linear2} \\
\rho\partial_t v_y = \eta (\partial_z^2-q^2) v_y-iq_y \Pi
-iK(\partial_z^2-q^2) q_x n_y \, ,
\label{linear3} \\
\rho\partial_t v_z=\eta(\partial_z^2-q^2)v_z
-\partial_z \Pi -iK(\partial_z^2-q^2)q_x n_z
-{\epsilon_\perp} \epsilon_0 E(t)(\partial_z^2-q^2)\Phi \, ,
\label{linear4}\\
 \partial_t [{\epsilon}_\parallel \epsilon_0 q_x^2 \Phi
+{\epsilon}_\perp \epsilon_0 (q_y^2-\partial_z^2)\Phi
-iq_x( \Delta {\epsilon} \epsilon_0 n_z E(t)
+i\zeta q_y n_y+\zeta \partial_z n_z)]
\nonumber \\
=-\sigma_\parallel q_x^2 \Phi -\sigma_\perp(q_y^2-\partial_z^2)\Phi -
i\Delta\sigma E(t) q_x n_z .
\label{linear5}
\end{eqnarray}
Here $q^2=q_x^2+q_y^2$, $n_y,n_z$ are the components of the director $\bm n$, describing its deviations from the equilibrium orientation, and $v_y,v_z$ are the components of the velocity. We introduce also $\delta \bm E=-\nabla \Phi$ as a perturbation of the homogeneous electric field. The parameter $\Pi$, figuring in the equations (\ref{linear3},\ref{linear4}), is the effective pressure, satisfying the relation
\begin{equation}
(\partial_z^2-q^2) \Pi = i\zeta E(t) q_x^3 n_z +\zeta q_x^4\Phi
+K(\partial_z^2-q^2) q_x(q_y n_y-i\partial_z n_z)
-{\epsilon}_\perp \epsilon_0 E(t) \partial_z (\partial_z^2-q^2) \Phi \ .
\label{gpressure}
\end{equation}
The condition (\ref{gpressure}) is a consequence of the incompressibility condition $iq_x v_x+iq_y v_y +\partial_z v_z=0$. The same condition enables one to exclude $v_x$ from the set of the dynamic variables.

One has to solve the set of the linear equations (\ref{linear1}-\ref{gpressure}) supplemented by suitable boundary conditions. The natural boundary conditions for the fields $n_y,n_z,v_y,v_z,\Phi$ are zero Dirichlet boundary conditions. Physically they are related to fixing director at the surface by the strong anchoring energy, to non-slipping boundary conditions for the velocity and to fixing the electric potential at the boundaries of the plates. The last condition is provided by the conducting electrodes on the surfaces of the nematic film. The zero Dirichlet conditions have to be supplemented by an additional boundary condition $\partial_z v_z=0$, following from the incompressibility condition $iq_x v_x+i q_y v_y +\partial_z v_z=0$ and zero values of the components $v_x$, $v_y$ at the boundaries. Thus we arrive at the boundary conditions for the film of thickness $d$, imposed at $z=\pm d/2$:
\begin{eqnarray}
&& n_y(z=\pm d/2)=0\, , n_z(z=\pm d/2)=0\ , \Phi (z=\pm d/2)=0,
\nonumber \\
&& v_y(z=\pm d/2)=0\, , v_z(z=\pm d/2)=0\ , \partial_z v_z (z=\pm d/2)=0.
\label{bc1}
\end{eqnarray}
From the formal point of view, the additional condition $\partial_z v_z=0$ is needed to specify the generalized pressure $\Pi$, see Eq. (\ref{gpressure}).

The system of equations (\ref{linear1}-\ref{linear5}) describes a set of modes, possessing a complicated temporal behavior in the external alternating electric field. One can extract eigen-modes, that are characterized by returning to their spacial structure after a period $T$ of the external field up to a factor $\Lambda_i$, that can be called the amplification factor of the eigen-mode. We introduce also the characteristic exponents $\lambda_i$ related to the amplification factors $\Lambda_i$ through $\Lambda_i=\exp(\lambda_i T)$. The exponents $\lambda_i$ determine the behavior of the eigen-modes on temporal scales much larger than $T$ via the factors $\exp(\lambda_i t)$. The equations (\ref{linear1}-\ref{gpressure}) written in terms of the variables $n_y,in_z,iv_y,v_z,\Phi,\Pi$ have real coefficients. That is why the characteristic exponents $\lambda_i$ must be all real or contain some complex conjugated pairs.

One easily checks that the set of the linear electro-hydrodynamic equations (\ref{linear1}-\ref{gpressure}) is invariant under the following transformations:
\begin{eqnarray}
q_y\to-q_y, \ n_z\to-n_z, \  n_y\to n_y,
\nonumber \\
v_z\to-v_z, \ v_y\to v_y,
\label{symmetry1}
\end{eqnarray}
and
\begin{eqnarray}
q_x \to -q_x, \ n_z\to n_z, \ n_y\to n_y,
\nonumber \\
v_z\to-v_z, \ v_y\to -v_y, \ \Pi\to -\Pi.
\label{symmetry2}
\end{eqnarray}
Note that even beyond the assumptions of our simplified model the complete set of the linear nemato-dynamic equations remains invariant under the transformations (\ref{symmetry1},\ref{symmetry2}). The symmetry of the equations under the transformations (\ref{symmetry1},\ref{symmetry2}) lead to the conclusion that the characteristic exponents $\lambda_i(q_x,q_y)$, do not change upon changing sign of $q_x$ or of $q_y$.

The modes described by the system of equations (\ref{linear1}-\ref{linear5}), all decay below the instability threshold. Above the threshold, a mode (or modes), that can be called the critical one, becomes unstable. Evidently, the actual patterns of the nematic above the threshold cannot be analyzed in the framework of the linear approximation (above the threshold the amplitude of the mode grows and the linear approximation is
violated). Nevertheless, some general conclusions concerning the state can be derived from the analysis of the system at the threshold.

Near the instability threshold the only third order terms in the dynamical equations should be taken into account, if the terms stabilize the solution of the equations. In the case the amplitude of the inhomogeneous contribution to the director field is small near the threshold. Then finally the only spatial harmonic survives with the wave vector $\bm q$, corresponding to the maximum value of $\mathrm{Re}\, \lambda_c(\bm q)$, where $\lambda_c$ is the characteristic exponent of the critical mode. Near the threshold $|\lambda_c|\ll \omega$ and, consequently, the slow dynamics of the director patterns is determined just by the critical characteristic exponent $\lambda_c$. If $\mathrm{Im}\, \lambda = 0$, then the pattern is stationary. Of course, the stationary behavior occurs on the time scale much larger than the period $T$ of the external alternating electric field, whereas oscillations of the pattern with the period $T$ are omnipresent.

Thus there are different possibilities, related to the value of the characteristic exponent $\lambda(q_x,q_y)$ of the critical mode. If $\mathrm{Im}\, \lambda = 0$ then the stationary patterns of the director field $\bm n$ are realized above the threshold, whereas the case $\mathrm{Im}\, \lambda \neq 0$ leads to the possibility of dynamic (oscillating in time) two-dimensional structure above the instability threshold. To avoid a misunderstanding, we stress again, that speaking about the stationary/oscillating regimes we imply time scales much larger than the period $T$ of the external alternating electric field.

If $\mathrm{Im}\, \lambda \neq 0$, then the critical mode oscillates with time. Thus we arrive at the traveling wave described by the factor $\exp(i \,\mathrm{Im}\lambda\, t + i\bm q \bm r)$. One might encounter two or four plain waves, depending on the arrangement of the values of critical wave vectors in $\bm q$-space. Then the nematic pattern appearing as a result of the instability consists of some traveling or standing waves, depending on conditions at the plates limiting the NLC sample. We are specially interested in the possibility and aim to establish conditions for its realization. Our detailed analysis based on the results of numerical computations is presented in subsequent sections.

Some words about the mechanism of appearing the oscillating critical mode. We assume that the conductivity of the nematic is relatively small, that is, it satisfies the inequality (\ref{frequencycond}). Then, as it follows from the equations (\ref{linear1}-\ref{linear5}), there is a slowly decaying "potential" mode with $\mathrm{Re}\, \lambda\, T \ll 1$. Indeed, in the limit $\sigma_\perp,\sigma_\parallel\to 0$ the quantity in the square brackets in the left hand side of Eq. (\ref{linear5}) is unchanged during the electric field period, that corresponds to the condition $\lambda=0$ for the mode. For finite $\sigma_\perp,\sigma_\parallel$ we obtain $|\lambda|\sim \sigma/(\epsilon_0 \epsilon_\perp)$, explaining the inequality $\mathrm{Re}\, \lambda\, T \ll 1$ for the "potential" mode, following from Eq. (\ref{frequencycond}).

The flexoelectric instability implies that the director dynamics is also a "soft" mode. At approaching to the threshold the value $\mathrm{Re}\, \lambda$ for the mode tends to zero. Therefore near the threshold there occurs a hybridization of the "director" and of the "potential" modes. The hybridization can lead to forming two mixed modes with complex conjugated characteristic exponents $\lambda$, then $\mathrm{Im}\, \lambda \neq 0$. At some conditions the mixed modes survive up to the instability threshold. Then $\lambda$ has a non-zero imaginary part at the threshold. We will examine the conditions needed for realization of this scenario.

\section{Flexoelectric instability in an unbounded nematic}
\label{sec:P1}

Here we consider the case of an unbounded nematic, enabling us to reduce the set of partial differential equations to the ordinary differential equations and to avoid difficulties related to account of the boundary conditions (\ref{bc1}). As we already noted, the approach allows one to scan quickly a wide range of material parameters to find the most interesting regions. If it is necessary, then the parameters can be examined in more detail for a nematic sample of finite thickness. The analysis of the film of the finite thickness is presented in the next Section \ref{sec:P2}.

For the unbounded nematic the set of equations (\ref{linear1}-\ref{linear5}) admits a solution, which is a linear combination of $\cos(q_z z)$ and $\sin(q_z z)$, where $q_z$ is an arbitrary parameter. Then the system of equations (\ref{linear1}-\ref{linear5}) are reduced to the ordinary differential equations for the coefficients at $\cos(q_z z)$ and $\sin(q_z z)$. In terms of the variables $n_y,in_z,iv_y,v_z,\Phi,\Pi$ the equations have real coefficients. That is why the amplification factors $\Lambda_i$ or the characteristic exponents $\lambda_i$ are either real or form the pairs of complex conjugated values. We fix $q_z$ and find $q_x,q_y$ corresponding to the critical mode, first achieving at the flexoelectric instability.

Alternatively one can look for a solution of the system (\ref{linear1}-\ref{linear5}) proportional to $\exp(iq_z z)$. Then the equations, following from Eqs. (\ref{linear1}-\ref{linear5}) for the unbounded case, are written as
\begin{eqnarray}
&&\partial_t n_y=i q_x v_y +\frac 1{\gamma}
\left[ -Kk^2 n_y -i \zeta q_y n_z E -\zeta q_xq_y\Phi \right] \ , \
\label{linear055} \\
&&\partial_t n_z =i q_x v_z +\frac 1{\gamma}
[ -K k^2 n_z +i \zeta q_y n_y E
+{\epsilon_0 \Delta \epsilon }E^2 n_z
\nonumber \\
&& -i {\epsilon_0 \Delta \epsilon }Eq_x \Phi
-\zeta q_x q_z \Phi ]\ ,
\label{linear155}\\
&&\rho\partial_t v_y = -\eta k^2 v_y-iq_y \Pi +iK k^2 q_x n_y \ , \
\label{linear255} \\
&&\rho\partial_t v_z=-\eta k^2 v_z-iq_z \Pi +iK k^2 q_x n_z
-{\epsilon_0 \epsilon}_\perp E k^2\Phi \
,\ \label{linear355} \\
&& -\partial_t [-{\epsilon_0  \epsilon}_\parallel q_x^2 \Phi
-{\epsilon_0  \epsilon}_\perp(q_y^2+q_z^2)\Phi
+iq_x( {\epsilon_0 \Delta \epsilon}\, n_z E+i\zeta q_y n_y+i\zeta q_z n_z)]
\nonumber \\
&&=-\sigma_\parallel q_x^2 \Phi -\sigma_\perp(q_y^2+q_z^2)\Phi +
i\Delta\sigma Eq_x n_z,
\label{linear55}
\end{eqnarray}
where $k^2=q^2+q_z^2=q_x^2+q_y^2+q_z^2$. The "pressure" $\Pi$ figuring in the equations is expressed as
\begin{equation}
-\Pi =i\zeta E q_x^3 n_z/k^2 +\zeta q_x^4\Phi/k^2
-K q_x(q_y n_y+q_z n_z)+i{\epsilon_0 \epsilon}_\perp E q_z \Phi \ ,
\label{unpressure}
\end{equation}
as a consequence of Eq. (\ref{gpressure}).

After substituting the expression (\ref{unpressure}) into the equations (\ref{linear055}-\ref{linear55}) they are reduced to the form
\begin{equation}
\frac{d \bm f}{dt} = \hat \Gamma \bm f,
\label{gammabm}
\end{equation}
where $\bm f$ is the vector with the components $n_y,n_z,v_y,v_z,\Phi$ and $\hat \Gamma$ is the matrix $5\times 5$ with components periodically varying as time $t$ goes.

As we noted, in NLCs the hydrodynamic flow degrees of freedom are much faster than the director mode. Therefore the velocity in the slow critical mode can be examined in the adiabatic approximation. That means that the time derivatives in the equations (\ref{linear255},\ref{linear355}) for $v_y$, $v_z$ can be neglected. Then the velocity components can be expressed from the expressions via $n_y, n_z, \Phi$. Substituting the expressions into Eqs. (\ref{linear055},\ref{linear155},\ref{linear55}) we find the equation of the form (\ref{gammabm}) for the three variables $n_y, n_z, \Phi$. Then we arrive at the matrix $\hat \Gamma$ $3\times 3$. We use both approaches, with the matrix $5\times 5$ and with the matrix $3\times 3$. The results, obtained in the framework of the approaches, appear to be the same.

One technical comment is in order here. The characteristic relaxation times of the essential dynamic modes (director, velocity, and potential modes) are strongly different. Therefore the computation of the evolution matrix $5\times5$ requires the solution of so-called rigid system of differential equations \cite{Press92}, that could create some problems for numerical computations. That is why we use both approaches, with the matrices $5\times 5$ and the matrices $3\times 3$, to be sure that the numerical results are correct.

We solve numerically the equation (\ref{gammabm}) on a period to find the evolution matrix $\hat W$:
\begin{equation}
\bm f(t+T)= \hat W \bm f(t).
\label{evolutionm}
\end{equation}
The evolution matrix $\hat W$ is independent on time $t$ thanks to periodicity of the matrix $\Gamma$. The eigen-values of the matrix $\hat W$ are no other than the amplification factors $\Lambda_i$ of the eigen-modes of the problem. Note that for $n$ periods
\begin{equation}
\bm f(t+nT)=( \hat W)^n \bm f(t).
\label{evolutionn}
\end{equation}
The eigen-values of the matrix $( \hat W)^n$ are $\Lambda_i^n$. Therefore, to better distinguish the critical mode, it is worth to examine the evolution determined by Eq. (\ref{gammabm}) on a few periods.

To find the evolution matrix $\hat W$ one can take as initial values $\bm f$ in Eq. (\ref{evolutionm}) the vectors $(1,0, 0,\dots)$, $(0,1,0,\dots)$ \dots. Then the corresponding vectors $\bm f(t+T)$ (found numerically) constitute the evolution matrix $\hat W$. The procedure can be conducted both for the matrices $5\times 5$ and $3\times$, and is directly generalized for the evolution on some periods. After computing the evolution matrix $\hat W$ we find its eigen-values $\Lambda_i$ and then the characteristic exponents from $\Lambda_i=\exp(\lambda_i T)$.

We are interested in the principal modes, i.e. the modes with maximal real part of the characteristic exponents $\lambda_i$, relevant for examining the flexoelectric instability. The oscillating in time regime of the critical mode is realized if there exist two modes characterized by complex conjugated values of the characteristic exponents with non-zero imaginary parts. Note that at $q_x=0$ the system of the equations (\ref{linear055}-\ref{linear155}) for the components $n_y,n_z$ of the director field is decoupled from the other equations. Then the evolution of $n_y,n_z$ is governed by an Hermitian matrix $2\times2$. Therefore the characteristic exponents $\lambda_i$ related to the director mode are real in this case. Since just the director mode is responsible for the flexoelectric instability, we conclude that the oscillating regime cannot be realized at $q_x=0$.

To simplify the dynamic analysis, one often neglects the hydrodynamic flow (see, e.g., \cite{EC19}). In the case we stay with the three variables, $n_y,n_z,\Phi$. Though the matrix, determining the time derivative of the variables is not Hermitian, we did not find a non-zero imaginary part of $\lambda_1$ in this case. Thus our results suggest that an account of the hydrodynamic degrees of freedom is crucial for achieving the regime with the oscillating in time critical mode.

\subsection{Pulsing external electric field}
\label{subsec:meander}

We analyze mainly the external harmonically varying electric field $E=E_0\cos(\omega t)$. To check whether the obtained results are robust we investigate also the pulsed form (\ref{meand}) of the alternating external electric field. The case enables one to examine the flexoelectric instability semi-analytically. As for the harmonically varying field, we study solutions of the system of the dynamic equations (\ref{linear055}-\ref{linear55}) during a period of the pulsing field. The evolution matrix $\hat W$ is collected from the solutions for the initial vectors $(1,0, 0,\dots)$, $(0,1,0,\dots)$ \dots as well.

For the pulsing field (\ref{meand}) the dynamic equations (\ref{linear055}-\ref{linear55}) during the time intervals $0<t<T/2$ and $T/2<t<T$ are the sets of linear differential equations with constant coefficients. Thus any solution of such system is a sum of the functions $\propto \exp (p\, t)$, where the set of the five exponents $p_\alpha$ can be obtained analytically. The temporal dependencies of the coefficients $n_y,n_z,v_y,v_z,\Phi$ during the first half-period are determined by the five initial conditions $(1,0, 0,\dots)$, $(0,1,0,\dots)$ \dots. To find the $t$-dependence of the coefficients $n_y,n_z,v_y,v_z,\Phi$ during the second half-period, one has to use the continuity conditions at $t=T/2$ for the variables $n_y,n_z, v_y,v_z$ and of the variable
\begin{equation}
-{\epsilon}_\parallel q_x^2 \Phi
-{\epsilon}_\perp(q_y^2+q_z^2)\Phi
+iq_x \Delta {\epsilon}\, n_z E,
\nonumber
\end{equation}
as it follows from Eqs. (\ref{linear055}-\ref{linear55}).

As a result of the procedure, we find $n_y,n_z,v_y,v_z,\Phi$ at $t=T$ for all five initial conditions. Thus the quantities, constituting the evolution matrix $\hat W$, can be found analytically. To find the eigen-values of $\hat W$ (amplification factors) $\Lambda_i$ one has to solve the characteristic equation $\mathrm{det}\, (\hat W -\Lambda)=0$. The equation is too complicated to be solved analytically. However, the equation can be easily solved numerically. This is the only numerical step needed to investigate the dynamic flexoelectric instability in the pulsating electric field for the unbounded nematic.

Our computations show that the results for the pulsing field agree well with the corresponding results for the harmonically varying external field. The threshold value of the amplitude of the pulsing field is approximately two times smaller that the amplitude of the harmonic field.

\section{Possible types of the instabilities}
\label{sec:transitions}

Here we illustrate different types of instability noted in Section \ref{sec:general} by presenting results of the numerical computations for suitable values of the material parameters. We draw the values of the amplification factor $\Lambda$ of the critical mode near the instability threshold as a function of the lateral wave vector $(q_x,q_y)$ at a given $q_z$. It is chosen to be less than the lateral wave vector.

Let start with discussing the case of stationary stripes. The case is most easily realized provided the instability occurs at $q_x=0, q_y\neq 0$. As we already noted, in the case the critical mode has real $\Lambda$. Therefore in this regime the stripe structure arises above the instability threshold with the stripes oriented along the $X$-axis. The case is realized if the dimensionless parameters (\ref{dparameters}) are relatively large.

\begin{figure*}
\hskip-0.3true cm
\includegraphics[scale=0.4]{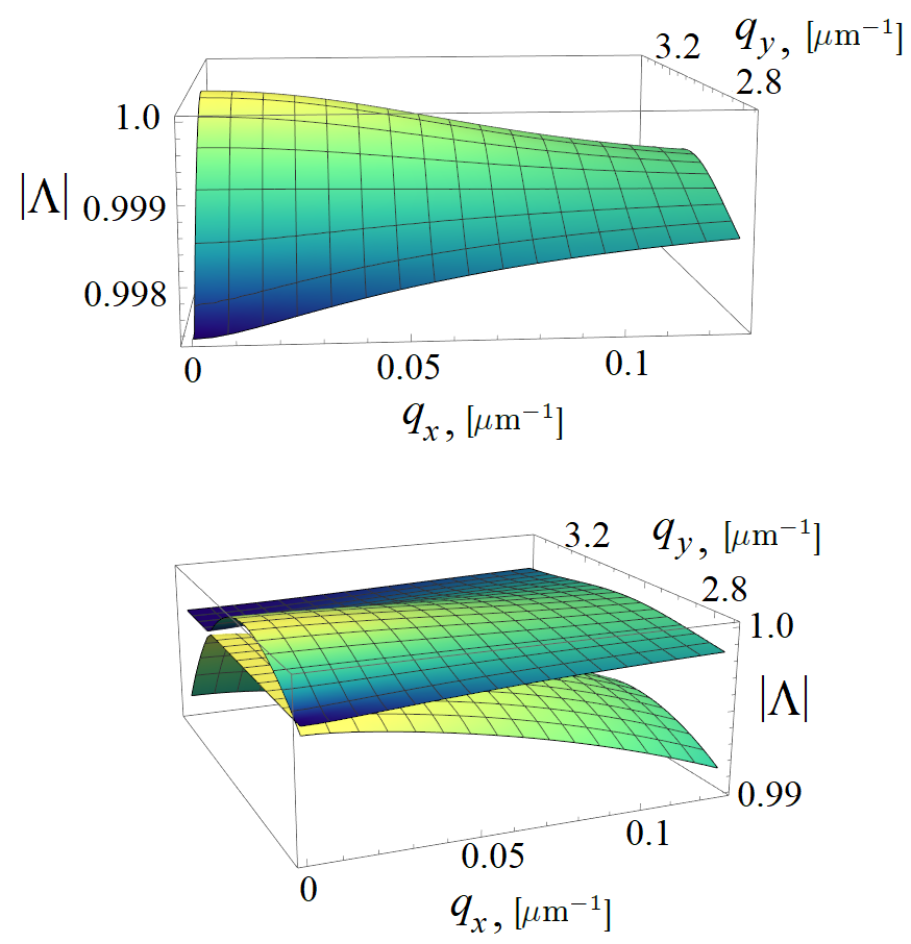} \label{triv1}
\caption{The dependence of the amplification factors $\Lambda $ of two principal modes
   on
$q_x$ and $q_y$  for $\omega= 2\pi \cdot 510 $ s$^{-1}$ :  $\Lambda = 1.00$  at  $E_{b c}= 3.73 \cdot 10^6$ V/m ,  $q_z = 0.04 \mu m^{-1}$, $\sigma_\perp = 1$ s$^{-1}=  10^{-10}$ $\Omega^{-1} \cdot m^{-1}$, $\Delta \sigma =  - 0.2 \sigma_\perp $,
   $K = 7 \cdot 10^{-12}\,$ N,
  $\zeta = 9.2 \cdot 10^{-4}\,$  SGSE \,units $= (9.2/3) \cdot 10^{-11}\,$ ${C}\cdot m^{-1}$,
  ${\epsilon}_\perp = 9.2 $,  $\Delta {\epsilon} = - {4\,\pi}\cdot 0.11$, $\gamma = \, 0.06\, Pa\cdot s\,$, $\eta = 0.06\, Pa\cdot s $.}
\label{Figure1}
\end{figure*}

To illustrate the statement we present our numeric results for the parameters
\begin{equation}
\eta/\gamma= 1 \, , \
\Delta\epsilon/\epsilon_\perp= -0.15 \,, \
K_1 |\Delta\epsilon| \epsilon_0/\zeta^2 = 0.091\, . \
\nonumber
\end{equation}
The complete set of the parameters used for the computations is presented in the caption to Fig. \ref{Figure1},
where the dependencies on $q_x$, $q_y$ of the amplification factors $\Lambda $ of two principal modes are presented near the instability threshold.

\begin{figure}
\begin{center}
\includegraphics[scale=0.45]{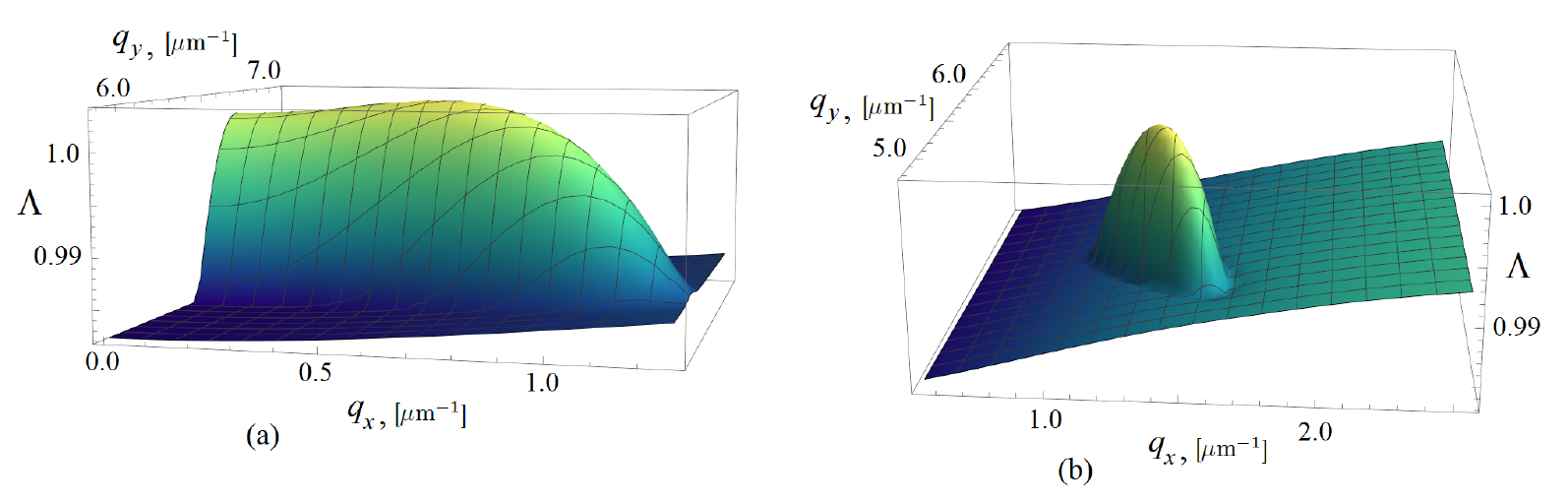} \label{nontrRa}
\end{center}
\caption{ The dependencies of the amplification factors $\Lambda_{1,2}$ of two principal modes on $q_x, q_y$ at $\omega= 2\pi \cdot 500 $ s$^{-1}$, $q_z = 0.41 \mu m^{-1}$, $K_1 = 7 \cdot 10^{-12}$ N, $K_2 = 5 \cdot 10^{-12}$ N, $K_3 = 5 \cdot 10^{-12}$ N,   ${\epsilon}_\perp = 14 $,  $\Delta {\epsilon} = - 3$, $\sigma_\perp = 10$ s$^{-1}= 10^{-9}$ $\Omega^{-1} \cdot m^{-1}$, $\Delta \sigma =  - 0.2 \sigma_\perp $, , $\eta = 0.066\, Pa\cdot s \,$
 (a)  :  $\Lambda = 1.0038$  at  $E_{b c}= 1.86\cdot10^6$ V/m for
   $\zeta = (15.16/3) \cdot 10^{-11}$ ${C}\cdot m^{-1}$,
     $\gamma = 0.054\, Pa\cdot s$;
      (b)  $\Lambda = 1.0006$  at  $E_{b c}= 1.46\cdot 10^6$ V/m for $\zeta = (18.5/3) \cdot 10^{-11}$ ${C}\cdot m^{-1}$,
            $\gamma = 0.035\, Pa\cdot s$.}
\label{difK}
\end{figure}

Besides the simple stripe structure some more complicated stationary structures can be realized above the instability threshold. They are realized if the instability is achieved at $q_x\neq 0, q_y\neq 0$. The possibility is discussed, e.g., in Refs. \cite{KP08,TE08,PK18}. In this case at the instability threshold there are four wave vectors $\pm q_x, \pm q_y$ corresponding to $\Lambda=1$. Above the instability threshold some stationary  periodic two-dimensional director pattern can occur. Alternatively, an oblique stripe structure can arise. The choice depends on the character of nonlinearity and is beyond the linear stability analysis.

To illustrate the possibility, we present numeric results for the parameters
\begin{equation}
\eta/\gamma= 0.82\, , \
\Delta\epsilon/\epsilon_\perp = - 0.214\, , \
K_1 |\Delta\epsilon| \epsilon_0/\zeta^2 = 0.073 \, , \
\nonumber
\end{equation}
in Fig. \ref{difK} (a),  and
\begin{equation}
\eta/\gamma= 0.53 \, , \
\Delta\epsilon/\epsilon_\perp = - 0.214 \, , \
K_1 |\Delta\epsilon| \epsilon_0/\zeta^2 =  0.049 \, , \
\nonumber
\end{equation}
in Fig. \ref{difK} (b). The complete set of the parameters used for the computations is presented in the caption to Fig. \ref{difK} (a) and (b).

The next case that can take place at the instability, is related to the oscillating director pattern above the instability threshold. It is realized if two critical modes with the complex conjugated amplification factors $\Lambda$ appear at the instability threshold, that is achieved at $q_x\neq 0, q_y\neq 0$. The case is realized at small enough parameters (\ref{dparameters}). Such situation probably may occur in nematic materials exhibiting phase transitions into various ferroelectric structures, see Refs. \cite{PI91,Kats2021,Nishikawa2017, Mertelj2018,Mertelj2022}.

\begin{figure*}
\includegraphics[scale=0.6]{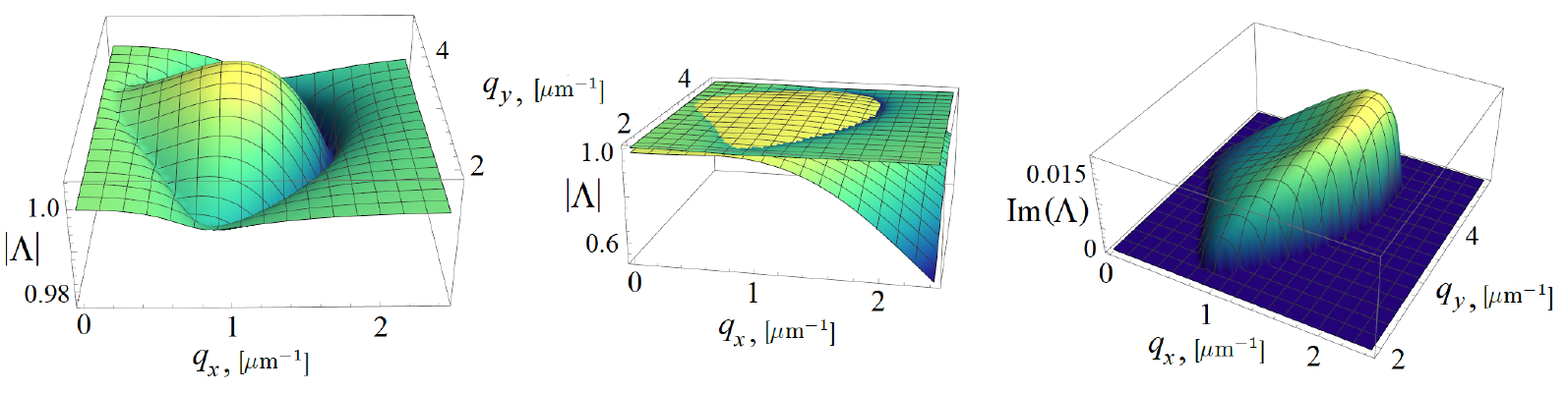}
\caption{Dependence of absolute values of amplification factor $\Lambda $ of two principal modes and imaginary part  of $\Lambda_1 $ on $q_x$ and $q_y$ for $\omega= 2\pi \cdot 500 $ s$^{-1}$ :  $\Lambda = 1.006 \pm\, i \,0.01$  at  $E_{b c}=  3.94 \cdot 10^6$ V/m for
$q_z = 0$, $\sigma_\perp = 1$ s$^{-1}=  10^{-10}$ $\Omega^{-1} \cdot m^{-1}$, $\Delta \sigma =  - 0.2 \sigma_\perp $,
   $K = 4 \cdot 10^{-12}\,$ N,
   $\zeta = (6.5/3)\cdot 10^{-11}\,$ ${C}\cdot m^{-1}$,
       ${\epsilon}_\perp = 22 $,  $\Delta {\epsilon} = - {4\,\pi}\cdot 10^{-1}$, $\gamma = \, 0.06\, Pa\cdot s\,$, $\eta = 0.06\, Pa\cdot s \,$. }
\label{comp1}
\end{figure*}

\begin{figure*}
\includegraphics[scale=0.6]{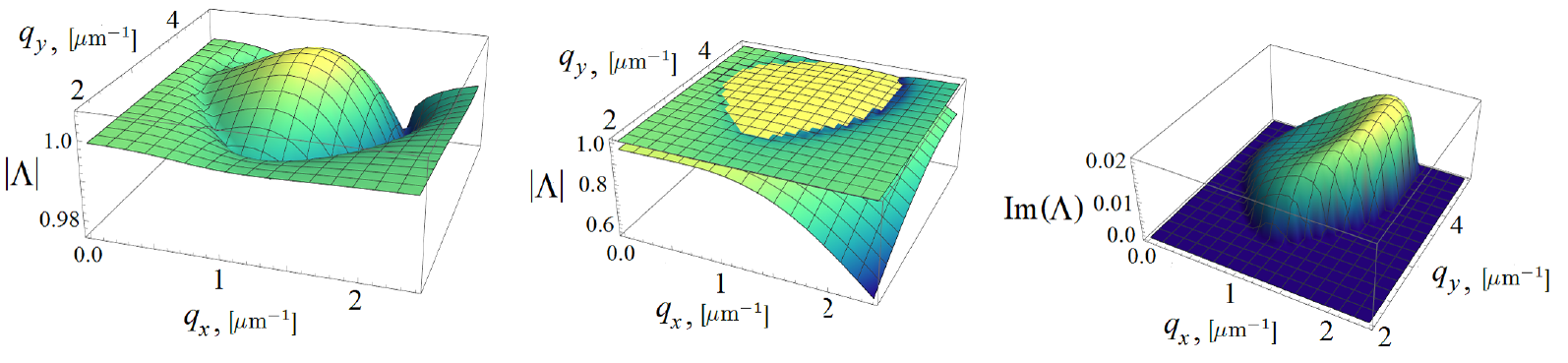}
\caption{The dependencies of absolute value of amplification factor $\Lambda $ of two principal modes and the imaginary part of $\Lambda_1 $ on $q_x$ and $q_y$ for $\omega= 2\pi \cdot 500 $ s$^{-1}$ :  $\Lambda = 1.0065 \pm i\, 0.013$  at  $E_{b c}=  3.94 \cdot 10^6$ V/m for
$q_z = 0.39\mu m^{-1}$, $\sigma_\perp = 1$ s$^{-1}=  10^{-10}$ $\Omega^{-1} \cdot m^{-1}$, $\Delta \sigma =  - 0.2 \sigma_\perp $,
   $K = 4 \cdot 10^{-12}\,$ N,
   $\zeta = (6.5/3)\cdot 10^{-11}\,$ ${C}\cdot m^{-1}$,
       ${\epsilon}_\perp = 24 $,  $\Delta {\epsilon} = - {4\,\pi}\cdot 10^{-1}$, $\gamma = \, 0.06\, Pa\cdot s\,$, $\eta = 0.06\, Pa\cdot s \,$. }
\label{osc1}
\end{figure*}

As an illustration of this possibility, we present numerical results for the following dimensionless parameters
\begin{equation}
\eta/\gamma= 1\, , \
\Delta\epsilon/\epsilon_\perp = - 0.057 \, , \
K_1 |\Delta\epsilon| \epsilon_0/\zeta^2 =  0.095 \,
\nonumber
\end{equation}
in Fig. \ref{comp1} for $q_z=0$ and for the parameters
\begin{equation}
\eta/\gamma= 1\, , \
\Delta\epsilon/\epsilon_\perp = - 0.052 \, , \
K_1 |\Delta\epsilon| \epsilon_0/\zeta^2 =  0.095 \,
\nonumber
\end{equation}
in Fig. \ref{osc1} for $q_z = 0.39\mu m^{-1}$, where the real and the imaginary parts of the complex amplification factor $\Lambda$ of the degenerated modes are presented. The complete set of parameters used for the simulation is presented in the caption to Fig. \ref{comp1} and  Fig. \ref{osc1}.

\begin{figure*}
\includegraphics[scale=0.5]{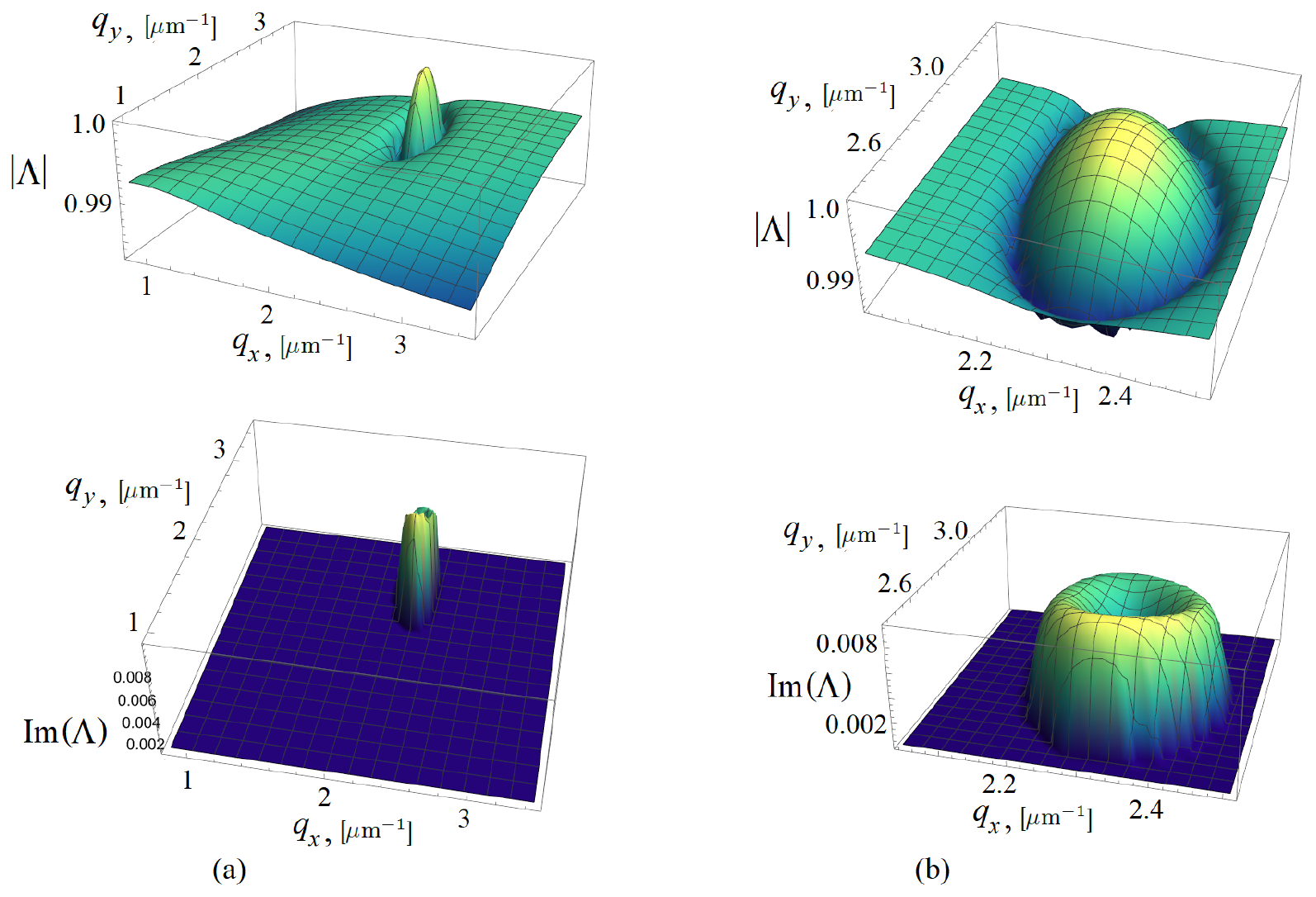}
\caption{ Dispersion laws near the threshold. The dependence of modules of amplification factor $\Lambda_1$ and imaginary part of $\Lambda_1$ of main mode on $\tilde q _x$ and $\tilde q _y$. $\Lambda = 1.00141  \pm i\, 0.001$  at  $E_{b c}= 0.65 \cdot 10^6$ V/m for
$q_z = 0.4 \mu m^{-1}$, $\sigma_\perp = 0.3$ s$^{-1}= 0.3\cdot 10^{-10}$ $\Omega^{-1} \cdot m^{-1}$, $\Delta \sigma =  - 0.7 \sigma_\perp $,
   $K_1 = 7 \cdot 10^{-12}$ N, $K_2 = 5 \cdot 10^{-12}$ N, $K_3 = 5 \cdot 10^{-12}$ N,
  $\zeta = (34/3)\cdot 10^{-11}$ ${C}\cdot m^{-1}$,  $e_3 = (34/3)\cdot 10^{-11}$ ${C}\cdot m^{-1}$,
     ${\epsilon}_\perp = 14 $,  $\Delta {\epsilon} = - 3$, $\gamma = 0.066\, Pa\cdot s$, $\eta = 0.015\, Pa\cdot s \,$. }
   \label{Ki1}
\end{figure*}

We checked that the oscillating regime can be realized for the case of different Frank modules and two different flexo-coefficients. The case is illustrated in Fig. \ref{Ki1}. The equations used for the computations can be found in Appendix \ref{app:finite}, Eqs. (\ref{1lvy})--(\ref{1lvz}). As it is seen from Fig. \ref{Ki1}, the absolute values of the amplification factor $\Lambda$ of two principal modes has two competing maxima, one of which is narrow and corresponds to complex $\Lambda$, whereas the second one is wide and corresponds to real $\Lambda$ (though with $q_x\neq 0$). Such possibility can be easily realized for the case of different Frank modules.

\subsection{Phase diagram}

One cannot study numerically the complete multidimensional phase space, and anyway it would be not very instructive. To gain a better understanding and some flavor of what can happen, it is beneficial to study the effects of the parameters chosen as selectively as possible. One of the way to explore the found above instabilities within the simplified model is to look for a plane of two parameters, the viscosity coefficient $\eta$ and the flexoelectric coefficient $\zeta$. The results are presented as "phase diagrams" plotted in terms of the dimensionless parameters $\eta/\gamma $ and $K\cdot \epsilon_0 |\Delta\epsilon |/\zeta ^2$. We put a series of the points to the diagrams, where the character of the critical mode is determined numerically. In the diagrams the star $*$ designates the critical phase with $q_x=0$ and real $\Lambda$, the open circle $\circ$ designates the phase with $q_x\neq 0$ and real $\Lambda$, and the bullet $\bullet$ designates the phase with $q_x\neq 0$ and complex $\Lambda$.

\begin{figure*}
\includegraphics[scale=0.55]{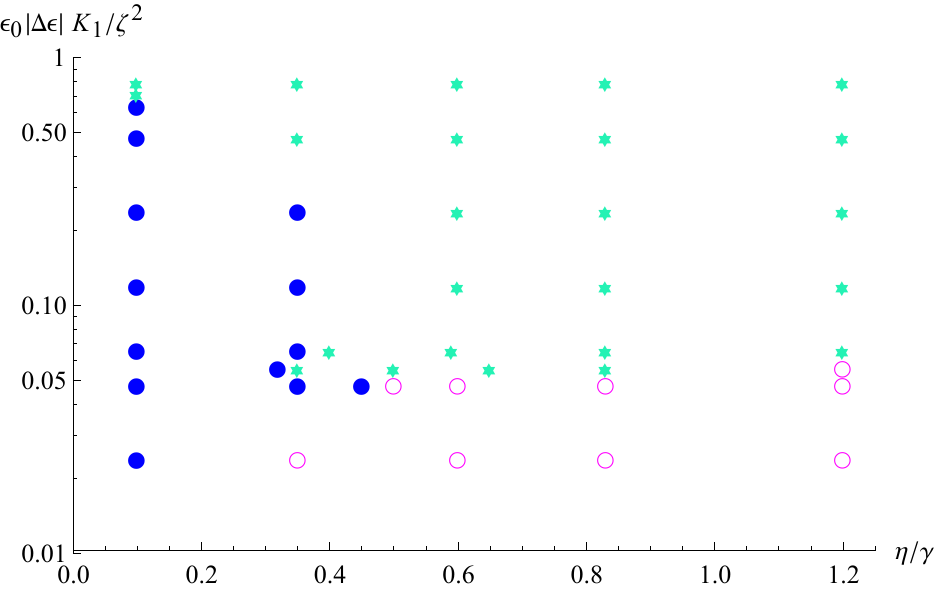}
\caption{The phase diagram of instabilities at different $\eta$ for the harmonic external field. Designations: $*$ is the phase with $q_x=0$ and real $\Lambda$, $\circ$ is the phase with $q_x\ne 0$ and real $\Lambda$, and $\bullet $ is the phase with $q_x\ne 0$ and complex $\Lambda$.}
\label{fig:pd1}
\end{figure*}

First, we take: $K_1=K_2=K_3=4\cdot 10^{-12}$ N, $\sigma_{\perp} =10$ s$^{-1}$, $\Delta\sigma =-2$ s$^{-1}$, $q_z=\pi/8~\mu m^{-1}$, $\epsilon_{\perp}=14$, $\Delta\epsilon =-0.8 \cdot \pi$, $\gamma =0.06~Pa\cdot s$, $\omega = 1000\cdot \pi$ s$^{-1}$. For the sinusoidal external field the results are presented in Fig. \ref{fig:pd1}, where we see all three possibilities. If both dimensionless parameters are large, the critical mode has $q_x=0$. At this condition the amplification factor $\Lambda $ is real. At the increase of $\zeta$, i.e. at the decrease of $K\cdot |\Delta\epsilon |/\zeta ^2$, the wave vector $q_x$ of the critical mode becomes $q_x \neq 0$ and the amplification factor $\Lambda $ remains real, if $\eta/\gamma$ is moderate. Finally at decrease of $\eta$ we pass to the critical mode with $q_x \ne 0$ and complex amplification factor $\Lambda $.

\begin{figure*}
\includegraphics[scale=0.55]{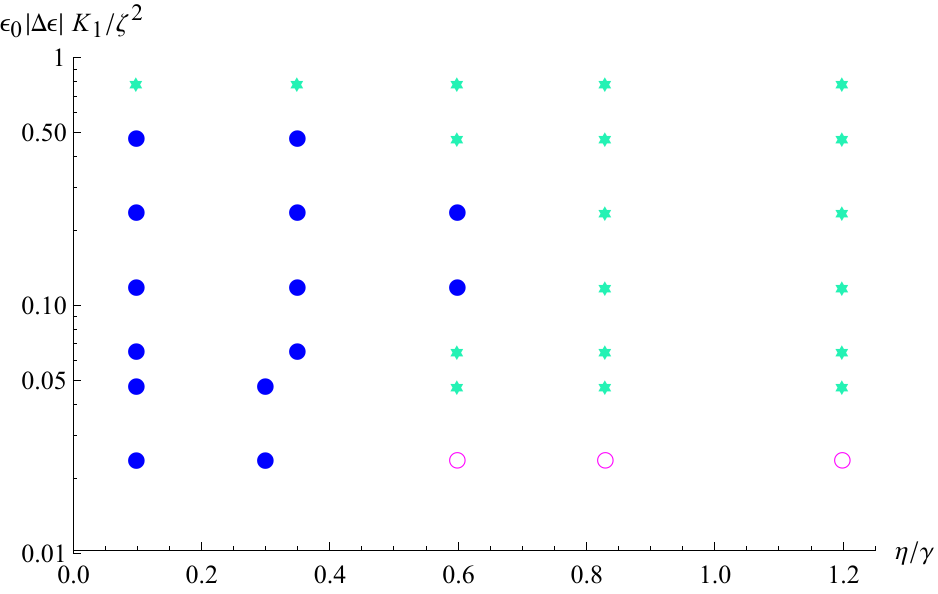}
\caption{The phase diagram of instabilities at different $\eta$ and $\zeta$ for the pulsing external field. Designations: $*$ is the phase with $q_x=0$ and real $\Lambda$, $\circ$ is the phase with $q_x\ne 0$ and real $\Lambda$, and $\bullet $ is the phase with $q_x\ne 0$ and complex $\Lambda$.}
\label{fig:pd2}
\end{figure*}

To check the "robustness" of the phase diagram presented in Fig. \ref{fig:pd1} we conduct an analogous investigation for the pulsing external field, see Eq. (\ref{meand}). The material parameters are the same as above. The results are presented in Fig. \ref{fig:pd2}. The general shape of this diagram is analogous to the diagram for the harmonic field, though the position of transition lines is different. For example, the transition to the phase where the critical mode with $q_x\neq 0$, real $\Lambda$ takes place at larger values of $\zeta $. The amplitude of the threshold external field for the pulsing field is smaller than for the harmonically varying field.

\begin{figure*}
\includegraphics[scale=0.55]{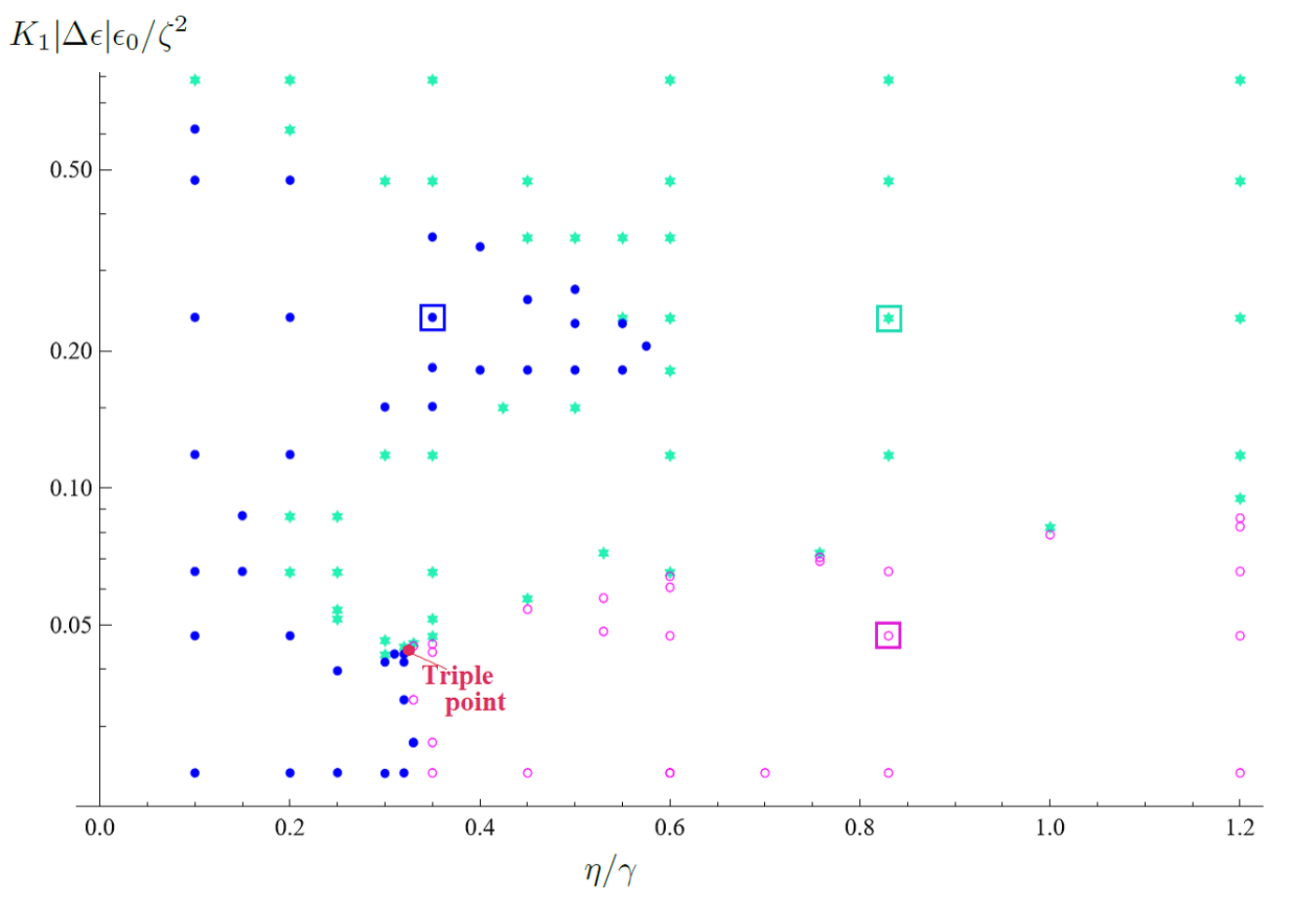}
\caption{The phase diagram of instabilities at different $\eta$ and $\zeta$ for harmonic external field and different Frank modules $K_1=7\cdot 10^{-12}$ N, $K_2=K_3=5\cdot 10^{-12}$ N, and under other parameters values: $q_z= 0.41 \mu m^{-1}$, $\sigma_\perp = 10$ s$^{-1}= 10^{-9}$ $\Omega^{-1} \cdot m^{-1}$, $\Delta\sigma = -  2$ s$^{-1}= 0.2 \cdot10^{-9}$ $\Omega^{-1} \cdot m^{-1}$, $\epsilon_{\perp}=14$, $\Delta\epsilon = - 3$, $\gamma =0.066 \, Pa\cdot s\,$, $\omega= 2\pi \cdot 500 $ s$^{-1}$.
 Designations: $*$ is the phase with $q_x=0$ and real $\Lambda$, $\circ$ is the phase with $q_x\ne 0$ and real $\Lambda$, and $\bullet $ is the phase with $q_x\ne 0$ and complex $\Lambda$.}
\label{fig:pd3}
\end{figure*}

The results presented above were obtained within our simplified model. The model is formulated with several assumptions (equal Frank constants, one flexoelectric coefficient, and a single viscosity coefficient), driven by pure desire to make formulas simpler. We do believe that more realistic description will not affect our qualitative conclusions, and transparency is worth a few oversimplifications. To illustrate our believe we compute the phase diagram for three different Frank modules, $K_1=7\cdot 10^{-12}$ N, $K_2=K_3=5\cdot 10^{-12}$ N. The results are presented in Fig. \ref{fig:pd3},  the complete set of parameters used for the computations is presented in the caption to the Figure. We see that the relative position of the three regions remains the same, though the borders between the phases are different. The results demonstrate universality of the phase diagram topology.

\subsection{Transitions between the regimes}

The aim of this subsection is to add a bit more details about transitions between the different regimes of the instability. On the phase diagrams the transitions are reflected as the borders between regions with different character of the instability.

\begin{figure}
\begin{center}
\includegraphics[scale=0.5]{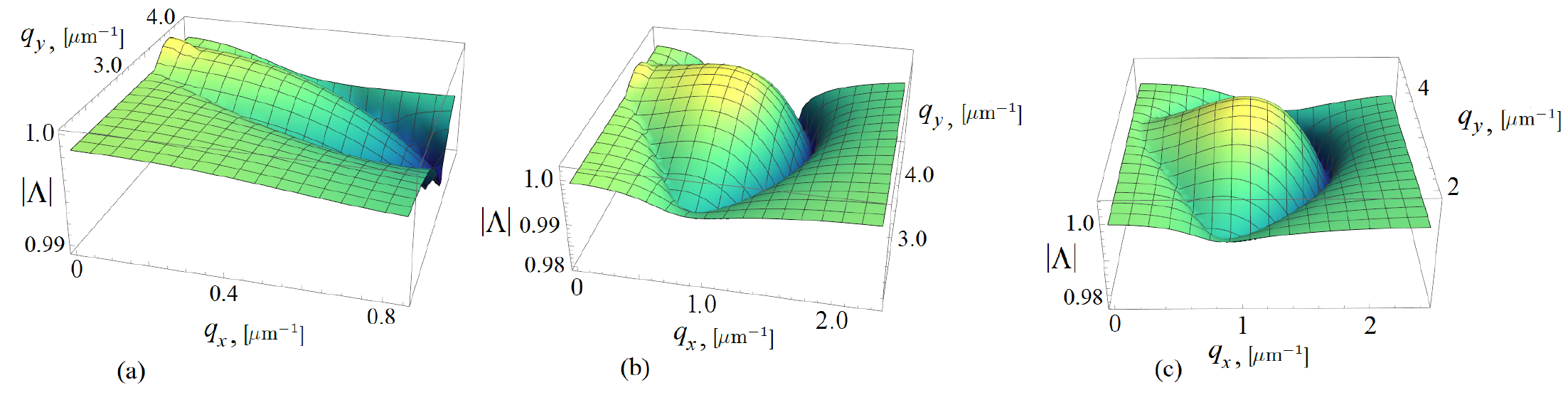}
\end{center}
\caption{Change of  principal modes spectrum $\vert \Lambda\vert$ under transition from static strips regime   to oscillating regime: (a)  $\epsilon_\perp = 16$ and $\Delta\epsilon/\epsilon_\perp=-0.079$; (b) $\epsilon_\perp = 20$ and $\Delta\epsilon/\epsilon_\perp-=0.063$; (c)$\epsilon_\perp = 22$  and $\Delta\epsilon/\epsilon_\perp=-0.057$ and other parameters: $\omega= 2\pi \cdot 500 $ s$^{-1}$,  $E_{b c}=  3.94 \cdot 10^6$ V/m,
$q_z = 0$, $\sigma_\perp = 1$ s$^{-1}=  10^{-10}$ $\Omega^{-1} \cdot m^{-1}$, $\Delta \sigma =  - 0.2 \sigma_\perp $,
   $K = 4 \cdot 10^{-12}\,$ N,
   $\zeta = (6.5/3)\cdot 10^{-11}\,$ ${C}\cdot m^{-1}$,
        $\Delta {\epsilon} = - {4\,\pi}\cdot 10^{-1}$, $\gamma = \, 0.06\, Pa\cdot s\,$, $\eta = 0.06\, Pa\cdot s \,$.}
\label{transav}
\end{figure}

\begin{figure*}
\includegraphics[scale=0.45]{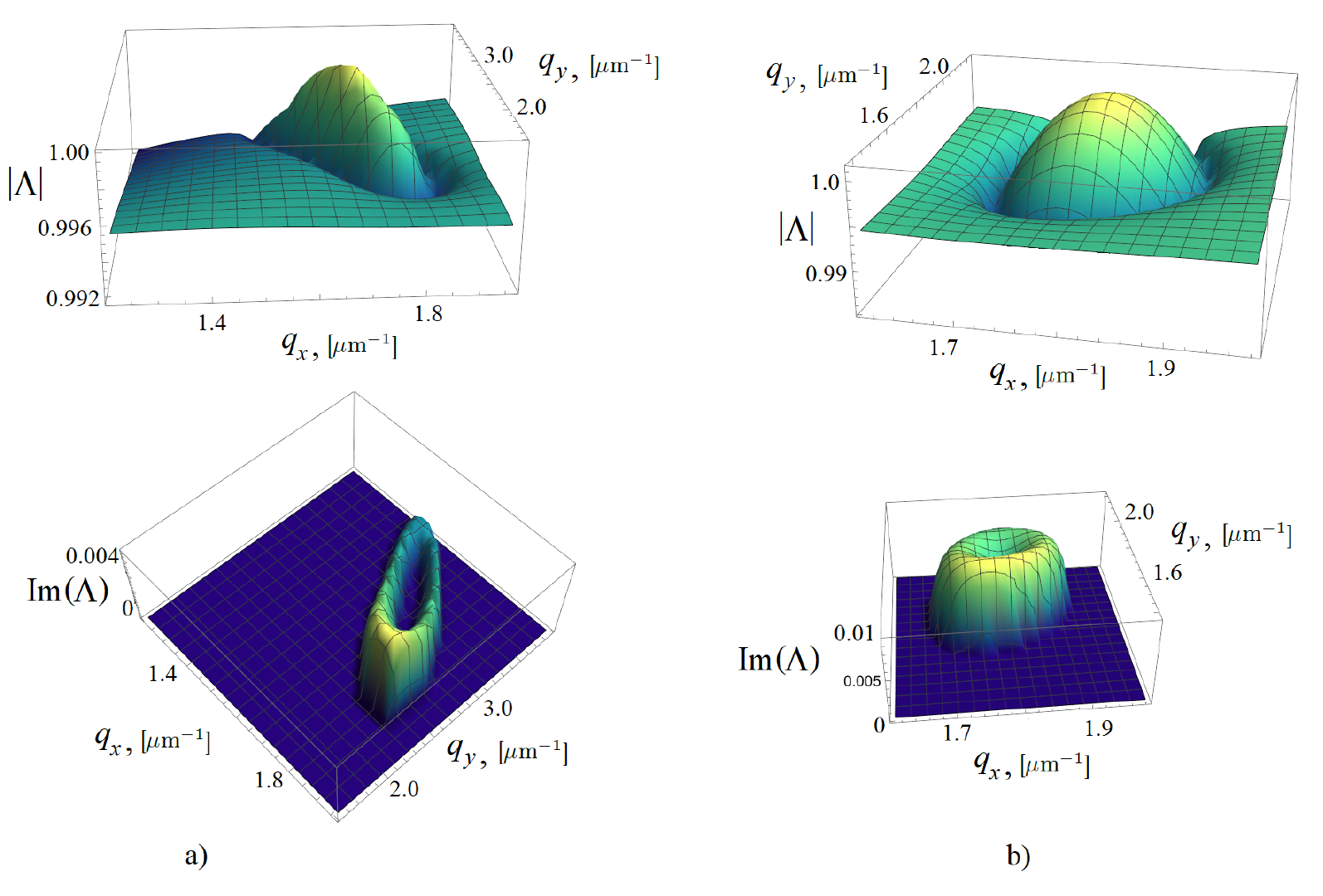}
\caption{Change of  principal modes spectrum $\vert \Lambda\vert$ under transition from  two-dimensional director pattern regime maximum with real $\Lambda$ to  oscillating regime: (a) $\gamma /\eta= 0.35/0.66$ and $E_{b c}= 0.93\cdot10^6$ V/m; (b) $\gamma /\eta= 0.25/0.66$ and $E_{b c}= 0.9\cdot10^6$ V/m;  and other parameters:
 $\omega= 2\pi \cdot 500 $ s$^{-1}$, $q_z = 0.4 \mu m^{-1}$, $\sigma_\perp = 10$ s$^{-1}= 10^{-9}$ $\Omega^{-1} \cdot m^{-1}$, $\Delta \sigma =  - 0.2 \sigma_\perp $,
   $K_1 = 7 \cdot 10^{-12}$ N, $K_2 = 5 \cdot 10^{-12}$ N, $K_3 = 5 \cdot 10^{-12}$ N,
   $\zeta = (26.46/3) \cdot 10^{-11}$ ${C}\cdot m^{-1}$,
      ${\epsilon}_\perp = 14 $,  $\Delta {\epsilon} = - 0.75$,  $\eta = 0.06\, Pa\cdot s$.}
 \label{transbv}
\end{figure*}

The transitions between the different regimes, which (depending of the material parameters) can be continuous (soft bifurcation) or jump-like (hard bifurcation). For example the transition  from the case $q_x=0$, real $\Lambda$ to the case $q_x\neq 0$, complex $\Lambda$ occurs discontinuously (with a jump from one type of the critical mode to another one),  see Fig. \ref{transav}. In turn, the transition from the case $q_x\neq 0$, real $\Lambda$ to the case $q_x\neq 0$, complex $\Lambda$ occurs continuously,  see Fig. \ref{transbv}.  In particular, Fig. \ref{transbv} demonstrates how the maximum with complex $\Lambda$ appears at the transition from two-dimensional pattern regime with real $\Lambda$, to the oscillating regime. The transition occurs continuously at the varying dimensionless parameter $\gamma /\eta$.

\begin{figure}
\begin{center}
\hskip-0.5true cm
\includegraphics[scale=0.65]{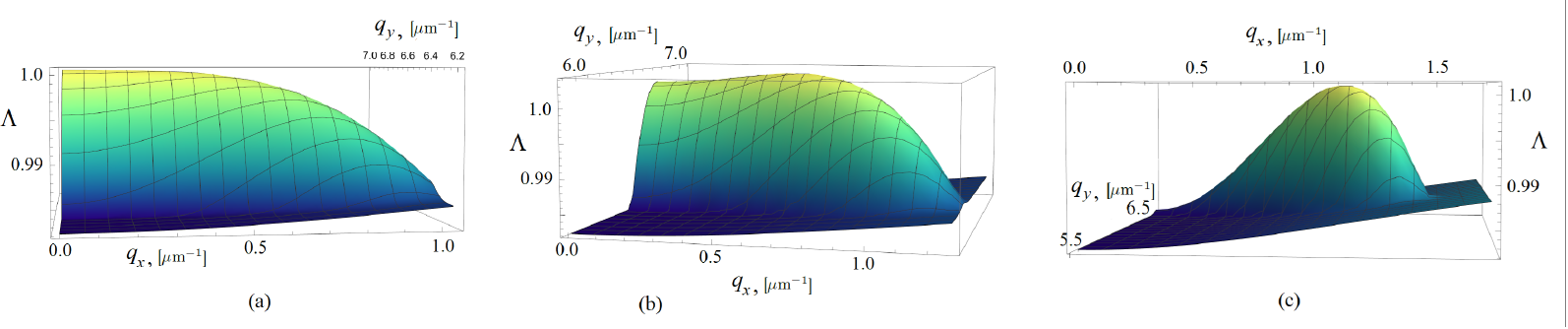}
\end{center}
\caption{Transition from the static stripes regime to two-dimensional director pattern one  with amplification factor maximum position: (a) $q_x = 0$, $q_y = 6.68 \,\mu m^{-1}$, $\eta/\gamma =  0.5/0.66=0.76$; (b) $q_x = 0.66\, \mu m^{-1}$, $q_y = 6.55 \,\mu m^{-1}$, $\eta/\gamma =  0.54/0.66=0.82$; (c) $q_x = 1  \mu m^{-1}$, $q_y = 6.3 \,\mu m^{-1}$, $\eta/\gamma = 1$.  Dependencies of  amplification factor $\Lambda_{}$ of the principal mode on $q_x, q_y$ at $\omega= 2\pi \cdot 500 $ s$^{-1}$,  $E_{b c}\simeq 1.86\cdot10^6$ V/m for
$q_z = 0.41 \mu m^{-1}$, $\sigma_\perp = 10$ s$^{-1}= 10^{-9}$ $\Omega^{-1} \cdot m^{-1}$, $\Delta \sigma =  - 0.2 \sigma_\perp $,
   $K_1 = 7 \cdot 10^{-12}$ N, $K_2 = 5 \cdot 10^{-12}$ N, $K_3 = 5 \cdot 10^{-12}$ N,
   $\zeta = (15.6/3) \cdot 10^{-11}$ ${C}\cdot m^{-1}$,
      ${\epsilon}_\perp = 14 $,  $\Delta {\epsilon} = - 3 $, $\gamma = 0.066\, Pa\cdot s$.}
\label{transba}
\end{figure}

As it concerns the transition from the case $q_x=0$, real $\Lambda$ to the case $q_x\neq 0$, real $\Lambda$, it can be either continuous or with a jump between the competing critical modes, depending on the material parameters. The continuous transition can be easily realized for the case of different Frank modules. This case is illustrated in Fig. \ref{transba}, corresponding to the continuous transition from the static stripes regime to the two-dimensional director pattern regime upon variation of the parameter $\eta/\gamma$.

It is instructive to examine in more details the transition to the oscillating regime from the stationary stripe structure. The transition occurs by the variations of the parameters listed in (\ref{dparameters}).
Near the transition a competition of potentially critical modes takes place. Typically, the modes with the complex amplification factor $\Lambda$, having maximum at $q_x\neq 0,q_y\neq 0$, compete with the mode with the real amplification factor $\Lambda$, which has the maximum at $q_x= 0,q_y\neq 0$. The transition takes place when the absolute values of the amplification factors of both modes are equal to unity, thus
it is a discontinuous transition. To illustrate the phenomenon, we present the computation results demonstrating the transition to the oscillating regime from the regime of stationary stripes under
\begin{equation}
\eta/\gamma= 1 \, , \
\Delta\epsilon/\epsilon_\perp = - 0.079 \div - 0.057 \, , \
K_1 |\Delta\epsilon| \epsilon_0/\zeta^2 = \, 0.095 \, , \
\nonumber
\end{equation}
in Fig. \ref{transav}. The transition occurs at varying (decrease) $\vert\Delta\epsilon\vert/\epsilon_\perp$. The complete set of the parameters used for the computations is presented in the caption to Fig. \ref{transav}.


Note that our calculations confirm that the instability (with the finite wave vectors) is also  possible only if $\zeta^2 > C\,\vert\Delta\epsilon\vert\,K\,$, with certain values of $C$ in accordance with the criterion (\ref{flexcrit}),  discussed  above. We also confirm  numerically, that the critical scaling condition (\ref{Ethr}), $\vert\Delta\epsilon\vert\, E_{c}^2 \, \propto \, \gamma \omega\,$, is valid.

\section{Film of finite thickness}
\label{sec:P2}

Here we consider the case where nematic is placed between two parallel plates. Computations of the dynamic flexoelectric instability for such case are more involved and computer time consuming than for an unbounded nematic. The complete system of the linear equations describing the nematic dynamics is the system of six equations (\ref{linear1}-\ref{gpressure}). All the equations are of the second order in $\partial_z$. In addition to the equations, one should use the boundary conditions (\ref{bc1}) at the surfaces of the film.

It is convenient to exclude the "pressure" $\Pi$ from the system of equations, thus reducing the number of the equations to five. Of course, the order of the equations after the exclusion is increased. It is possible to obtain an equation of the fourth order for $v_z$, keeping the orders of the equations for other variables to be equal to two. The system of such equations ideally corresponds to the boundary conditions (\ref{bc1}). Let us sketch a derivation of the equations.

To find the equation for $v_z$, one applies $\partial_z^2-q^2$ to Eq. (\ref{linear4}) and then expresses $(\partial_z^2-q^2) \Pi$ from Eq. (\ref{gpressure}) to obtain
\begin{eqnarray}
\rho(\partial_z^2-q^2)\partial_t v_z=\eta(\partial_z^2-q^2)^2v_z
\nonumber \\
-i\zeta E(t) q_x^3 \partial_zn_z -\zeta q_x^4\partial_z\Phi
-K(\partial_z^2-q^2) q_x(\partial_zq_y n_y-i q^2 n_z)
+{\epsilon}_\perp \epsilon_0 E(t) q^2 (\partial_z^2-q^2) \Phi
\label{film1}
\end{eqnarray}
Next, applying $\partial_z$ to Eq. (\ref{linear4}) and expressing then $\partial_z^2 \Pi$ from Eq. (\ref{gpressure}), one finds
\begin{eqnarray}
q^2 \Pi=-\rho\partial_t \partial_zv_z
+\eta(\partial_z^2-q^2)\partial_zv_z
\nonumber \\
 - i\zeta E q_x^3 n_z -\zeta q_x^4\Phi
-K(\partial_z^2-q^2) q_x q_y n_y
\label{film2}
\end{eqnarray}
Substituting the expression into Eq. (\ref{linear3}), we find the equation for $v_y$. There is the term proportional to $\partial_z^3 n_z$ in Eq. (\ref{film1}). Applying $\partial_z$ to Eq. (\ref{linear2}) we can then express $\partial_z^3 n_z$, then substituting it into Eq. (\ref{film1}). In more detail, the derivation and the resulting equations are presented in Appendix \ref{app:finite}, Eqs. (\ref{1lvy})--(\ref{1lvz}).

One can check, that after substitution (c.f. with \cite{KP08})
\begin{eqnarray}
v_x \, \to \,- i\,v_x \ ,
\nonumber \\
v_y \, \to \,- i\,v_y \ ,
\nonumber \\
n_z \, \to \,- i\,n_z \
\, , \
\label{subst0}
\end{eqnarray}
to the equations we arrive at the system of equations with real coefficients for the new variables.

In the framework of our computational scheme, we solve Cauchy problem for the system of differential equations for the fields $u_\alpha(z)=(n_y,n_z,v_y,v_z,Phi)$, starting from some initial condition at $t=0$ and satisfying the boundary conditions (\ref{bc1}). The problem is solved numerically for a period of the external electric field or for a number of periods.

To examine the instability, we use the scheme based on approximating the functions $u_\alpha$ by an expansion over a finite basis. Namely, we chose a set of $N_f$ basic functions $g_{\alpha,j}(z)$ where the first subscript, $\alpha=1\div 5$ corresponds to the fields $n_y,n_z,v_y,v_z,\Phi$ and the second subscript $j=1\div N_f$ numerates the functions of the set. Thus we arrive at the expansion
\begin{equation}
u_\alpha(z)=\sum_{j=1}^{N_f}
f_{\alpha,j} g_{\alpha,j}(z),
\label{expanu}
\end{equation}
where $f_{\alpha,j}$ are some coefficients.

To guarantee the correct boundary conditions for the fields $n_y,n_z,v_y,v_z,\Phi$ the appropriate boundary conditions should be imposed on the basis functions $g_{\alpha,j}(z)$. Namely, the functions $g_{\alpha,j}$ should be zero at $z=\pm d/2$. In addition, the functions $g_{4,j}$, figuring in the expansion of $v_z$, should have zero $z$-derivatives at $z=\pm d/2$, see Eq. (\ref{bc1}). We use the following basic set of the functions. For the fields $n_y$, $n_z$,  $v_y$, and $\Phi $ we use the sinusoidal basic functions
\begin{equation}
g_{\alpha,j}(z) = \sin [\pi j (z/d+1/2)], \quad  \alpha=1,2,3,5, \quad  j=1,\ldots N_f;
\label{i01}
\end{equation}
equal to zero at $z=\pm d/2$. For the field $v_z$ we choose the basic functions
\begin{equation}
g_{4,j}(z)= \,(z^2-d^2/4)\sin [\pi j (z/d+1/2)], \quad  j=1,\ldots N_f;
\label{i02}
\end{equation}
satisfying both boundary conditions, $v_z=0$ and $\partial_z v_z=0$.

To pass from the fields $u_\alpha(z)$ to the coefficients $f_{\alpha,j}$ one should define the projection procedure $u\to f$. For the purpose we use the metrics $L_2$, then
\begin{equation}
f_{\alpha,a} =\sum_b M^{-1}_{\alpha,ab} \int dz\, g_{\alpha,b}(z)\, u_\alpha(z) \, . \
\label{proj1}
\end{equation}
The matrices $\hat M_{\alpha}$ in Eq. (\ref{proj1}) are $5$ matrices $N_f \times N_f$ with the components
\begin{equation}
M_{\alpha,ab}=\int dz\, g_{\alpha,a}(z)\, g_{\alpha,b}(z)\, . \
\label{proj2}
\end{equation}
The matrices $\hat M_{\alpha}$ are, obviously, symmetric.

After solving Cauchy problem on one period and projecting the initial and the final values of the fields $u_\alpha$ in accordance with Eq. (\ref{proj1}), we find the coefficients $f_{\alpha,j}(0)$ and $f_{\alpha,j}(T)$. The linear character of our problem means that
\begin{equation}
f_{\alpha,a}(T) = \sum_{\beta,b} W_{\alpha,a;\beta,b}\,
f_{\beta,b}(0) \, . \
\label{evolfield}
\end{equation}
The matrix  $\hat W$ is $5N_f\times 5N_f$ matrix, which  represents  a generalization of the matrix $5\times 5$ figuring in Eq. (\ref{evolutionm}) for the unbounded case. We call the matrix  $\hat W$ the evolution matrix as well. Eigen-values $\Lambda_i$ of the evolution matrix $\hat W$ determine the amplification factors of the eigen-modes during the period $T$ and, consequently, the characteristic exponents via the relations $\Lambda_i=\exp(\lambda_i T)$. Since after the substitution (\ref{subst0}) we deal with the differential equations with real coefficients, the amplification factors $\Lambda_i$ or the characteristic exponents $\lambda_i$ are all real or, in addition to the real parameters, there are pairs of complex conjugated parameters.

The evolution matrix $\hat W$ can be collected, if we subsequently solve Cauchy problem for the set of the initial functions $g_{\alpha,a}(z)$ at $t=0$. Expanding the resulting functions at $t=T$ in accordance with the rule (\ref{proj1}), we find the set of the coefficients $W_{\alpha,a;\beta,b}$ constituting the evolution matrix $\hat W$, determining the evolution of the system during the period. Ultimately, we are interested in
the eigen-functions of the evolution matrix $\hat W$ and in the corresponding eigen-values $\Lambda$. More precisely, we are interested in the eigen-function with the principal $\Lambda$ (that is with maximal $|\Lambda|$), since just the eigen-function describes the instability.

Luckily it turns out that for the critical mode the amplitudes at the basis functions $g_j$ in the eigen-function expansion decay sufficiently fast with its number with the use of the reasonable set of the basis functions. The values of the amplification factors $\Lambda$ obtained for different numbers $N_f$ of the basis functions converge as $N_f$ increases. This fact verifies the regular convergence of the computational procedure. Thus, the described procedure allows one to obtain rather accurate quantitative description of the modes for the finite thickness nematic film.

We established that with the accuracy, needed for us, it is enough to take six basic functions, $N_f=6$. Then the matrix $\hat W$ is the matrix $30 \times 30$, where $30$ is the product of the number of the fields (five) and the number of basic functions ($N_f=6$). It is not very easy to compute all eigen-values of the matrix $30 \times 30$. However, we are interested solely in the principal mode. Therefore one can take the evolution during many (say $N$) periods. Then the eigen-values of the corresponding evolution matrix are $\Lambda^N$. This trick essentially simplifies investigation of the principal mode.

\begin{figure*}
\includegraphics[scale=0.2]{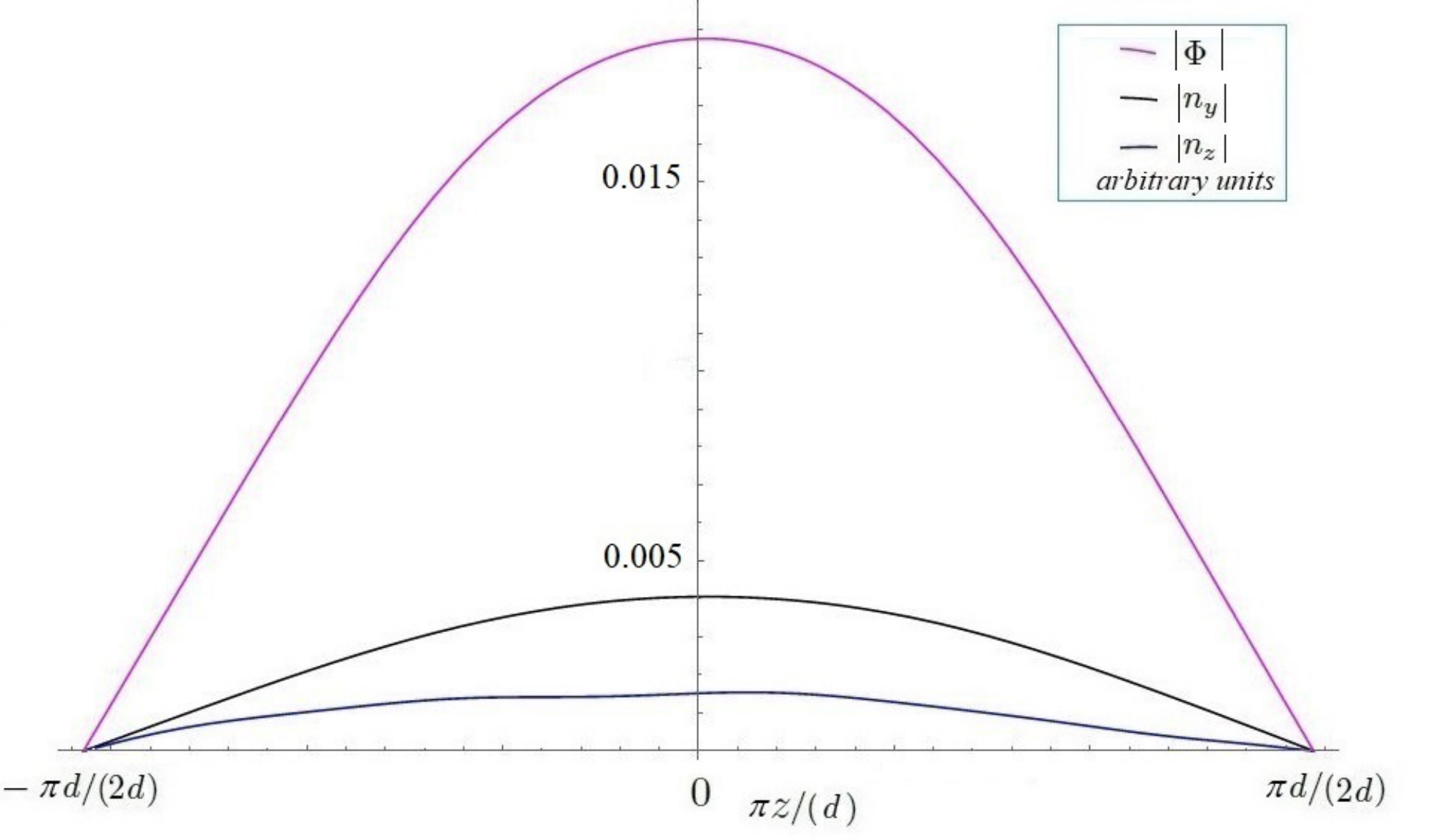}
\includegraphics[scale=0.2]{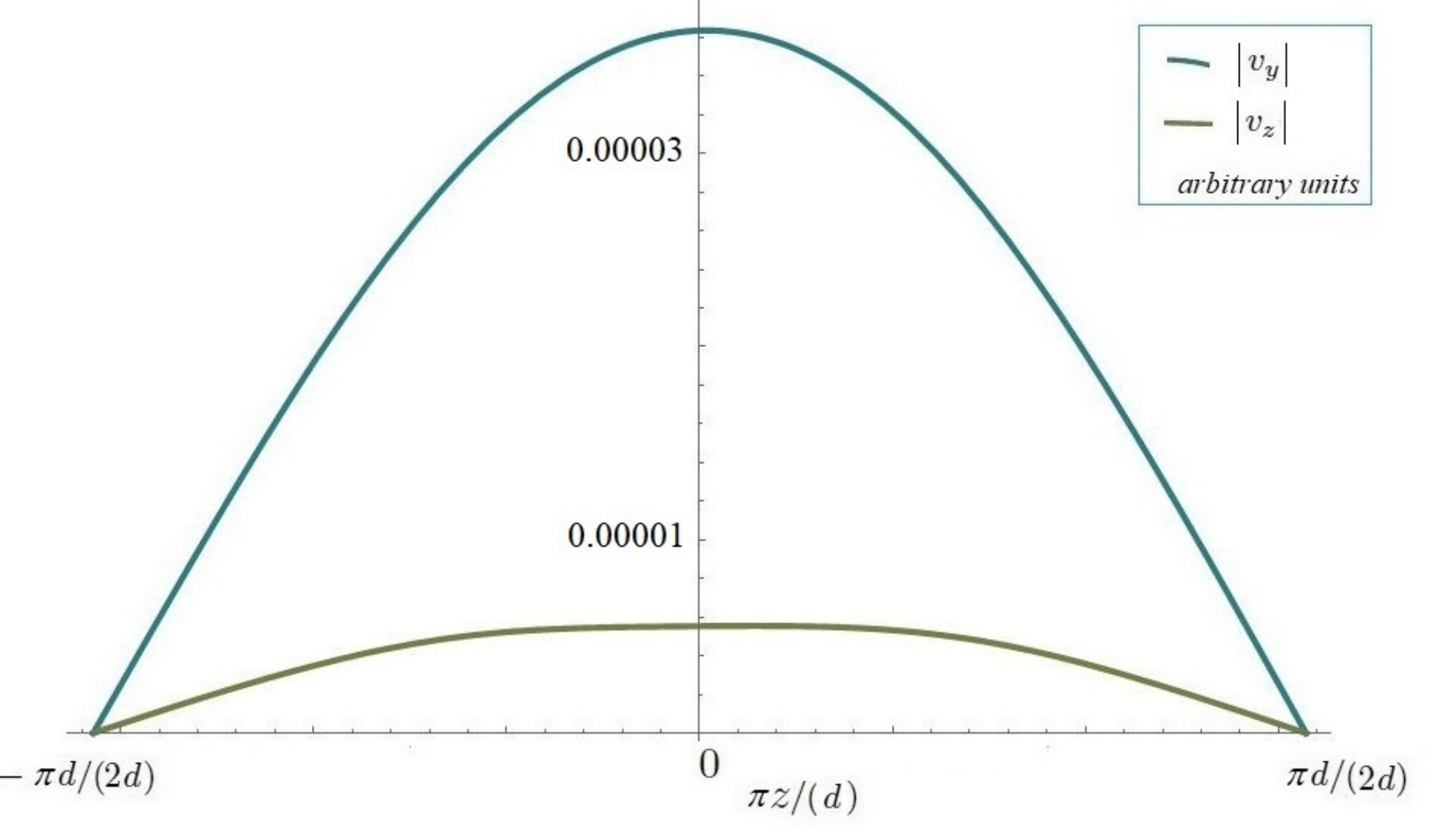}
\caption{Principal eigen-modes for the fields $n_y, n_z, \Phi $ and for the fields $v_y, v_z$ as functions of $\pi z/d$, for $t=150 T$.}
\label{modes1}
\end{figure*}

The principal eigen-modes near the threshold found as a result of the described procedure for a certain set of the material parameters are plotted in Fig. \ref{modes1}. The behavior of the eigen-functions is fairly smooth. One can see from Fig. \ref{modes1}, that the eigen-modes are close to $\sin[\pi(z+d/2)/d]$ (that is to the behavior characteristic for the unbounded case) everywhere except for the narrow regions near the boundary. That confirms effectiveness of the simplified analysis in the framework of the unbounded medium. The reason for this is smallness of $\pi/d$ in comparison with $q_y$ of the critical mode.

To justify explicitly that the results for the unbounded nematic and for the film of finite thickness are close, we have performed the computations for both cases with the identical sets of the material parameters. As an illustration, we have chosen three sets of the material parameters corresponding to all three types of the flexoelectric instability.  They correspond to the points marked by squares on the phase diagram in Fig. \ref{fig:pd3}. The results are summarized in the tables given in Appendix \ref{app:comparison}. In the tables we show the values of the critical electric field and the values of the amplification factor $\Lambda$ in the vicinity of the critical wave vector. The results confirm our expectations.

Note that when we investigate phase behavior of the system over varying film thickness $d$ only, under fixed other material parameters, starting from oscillating regime, we always go upon decreasing of $d$ to stationary stripe regime (trivial maximum with $q_x = 0$).

\section{Conclusion}
\label{sec:conclu}

In this work we find that the flexoelectric instability of NLCs in the external alternating electric field can lead to different inhomogeneous spatial-temporal structures of the director field $\bm n$ if a certain threshold value of the field is exceeded. The subject of our consideration are nematics with negative anisotropy of both, the dielectric permittivity $\epsilon_\parallel-\epsilon_\perp<0$ and the conductivity $\sigma_\parallel -\sigma_\perp<0$, where as above the subscripts $\parallel$ and $\perp$ designate the components along the director $\bm n$ and perpendicular to it. We analyzed the nematic film placed between two parallel conducting plates with the alternating electric potential differences applied to the plates. Then the electric field $\bm E$, directed perpendicular to the plates, is induced inside the film. Below the transition the state is assumed to be homogeneous with the director oriented along the surfaces of the plates. Such ordering is caused by suitable boundary conditions for the director at the surfaces.

We study numerically the flexoelectric instability based on the linearized dynamic equations of nematics.
The set of the equations cannot be solved analytically in the external alternating electric field. To find the qualitative peculiarities of the flexoelectric instability, we studied first the case when the boundary conditions can be ignored. Therefore after Fourier transform all the fields have harmonic dependence with a wave vector $q_z$ in the direction perpendicular to the plates. Then the equations can be easily solved numerically. To check whether such approach is physically adequate, we performed much more involved computations for the nematic films of finite thickness for the planar boundary conditions for the director, no slipping boundary conditions for the hydrodynamic velocity and fixing the electric potential at the plates. We show that the results for  both cases are close to each other if we choose $q_z=\pi/d$, where $d$ is the thickness of the film. This a bit surprising finding can be crudely understood by the fact that for Fourier harmonics relevant for the instability characteristic space scale along the $z$ axis is smaller than that in the orthogonal plane.

The flexoelectric instability is related to the distortions of the director field. However, in dynamics the director is coupled to hydrodynamics and electromagnetic degrees of freedom, that leads to a complicated structure of the critical mode, becoming unstable at increasing the external electric field. If the harmonic with $q_x=0$ and $q_y\neq 0$ becomes unstable first then a stationary stripe structure appears above the threshold. This scenario is well-known and described in the literature. Alternatively, the projection of the wave vector to the film plane is tilted to the equilibrium orientation of the director. Such possibility was discussed in Refs. \cite{KN04,KP08,TE08,PK18}. In this case the actual structure of the state above the threshold is determined by non-linear terms. Both a tilted stripe structure or a two-dimensional stationary periodic in space structure could appear above the threshold.

We revealed the third possibility (somehow overlooked in the previous works), the mode with the tilted wave vector and with the complex characteristic exponent of the critical mode. Then a non-stationary (oscillating in time) structures appear above the threshold. They are propagating or standing waves. As far as we know such structures were not discussed or investigated so far. We demonstrate numerically that the scenario with the oscillating structures can take place at a set of parameters for the computations of the unbounded nematic.

We show that the third scenario is realized at the condition (\ref{frequencycond}), leading to the existence of a slowly decaying "potential" mode describing the relaxation of the electric field fluctuations. Then in a range of the wave vectors $q_x$ and $q_y$ the hybridization occurs of the "potential" mode and of the "soft director" mode, related to the director instability. The hybridization could lead to the instability characterizing by a complex characteristic exponent $\lambda$, determining the behavior $\propto \exp(\lambda t)$ of the critical mode on times larger than the period. In the case $\lambda$ is purely imaginary at the threshold. Note that the scenario needs that the both components of the lateral wave vector of the critical mode be nonzero ($q_x \neq 0$, and $q_y \neq 0$).

The condition (\ref{frequencycond}) is crucial for our scenario. It can be achieved e.g., in purified from impurities NLCs (to the point see discussion of this issue in Refs. \cite{lavrentovich1,lavrentovich2,lavrentovich3}). Let us stress that the obligatory ingredient for the third scenario is the coupling of director to hydrodynamic degrees of freedom (neglecting hydrodynamic velocity ${\bm v}$ and its fluctuations makes the third scenario impossible). Our computations show also that the dielectric anisotropy of the nematic should be relatively weak for the realization of the third scenario. The condition can be satisfied for NLCs near the transition temperature to the isotropic phase (or for the alternating field frequency in the vicinity of the so-called inversion point, where the dielectric anisotropy changes its sign).

It is worth to noting that upon decreasing the film thickness $d$ (keeping all other parameters fixed), we always find the transition from the oscillating in time two dimensional patterns into the stationary stripe regime (with $q_x = 0$). The oscillating in time patterns can be realized for the relatively thick films. Quantitatively, it implies the condition $q d \gtrsim 1$, where the characteristic wave vector $q$ of the unstable harmonic is determined by Eq. (\ref{escrit2}).

We examined mainly the case of the external electric field harmonically varying with time. One can consider another possibility of the external alternating electric field. Namely, the field can be the pulse function (telegraph process). Then the system of equations for the unbounded nematic can be solved practically up to the end without numerical computations. We show that the results derived in the framework of this model are similar to those for the sinusoidal field. Particularly, the oscillating in time or propagating patterns are realized for some region of parameters near the threshold field. The results confirm the universal character of our findings.

We should admit that there are some physical ingredients missed in our approach. Note, as an example, the external, injected from electrodes charges, leading to the local violation of the electroneutrality \cite{EC19}. Note to the point that a finite electric current limits the overall thermalization of the NLCs, and in such that the instantaneous values of the material parameters can additionally be position dependent across the sample. Another missing in our publication ingredient is possible non-uniformity of the director surface anchoring. This ingredient has been introduced recently \cite{abbott1,abbott2} to simulate numerically localized and propagating excitations in the NLCs. However in this works the hydrodynamic degrees of freedom and finite conductivity were disregarded. Keeping everything said above in mind, we should admit that further theoretical and experimental work is required before a full understanding of dynamical flexoelectric instability in NLCs is reached. Nevertheless we do believe, that the physics behind our simplified model has to be understood before adding the additional ingredients. The same concerns non-linear effects. The solution of the non-linear dynamic equations is needed to identify the structure above the threshold. The analysis is a subject of future works.

\acknowledgments

The work of E.S.P. and V.V.L. was supported by the Russian Science Foundation (Grant No. 23-72-30006), and their work connected with derivation of linear equations was supported by the State assignment No. FFWR-2024-0014 of the Landau Institute for Theoretical Physics of the RAS. The work of E.I.K.  was supported by the State assignment No. FFWR-2024-0014 of the Landau Institute for Theoretical Physics of the RAS. The work of A.R.M. was supported by the State Assignment FMME-2022-0008  (N. 122022800364-6).

\medskip

The authors have no conflicts to disclose.

\appendix

\section{Derivation of linear equations}
\label{sec:LinEqs}

Let us linearize the full set of the nonlinear dynamic equations \cite{PM23} for NLCs in the a.c, external electric field. For small perturbations of the uniform initial  state with the electric field $E(t)\equiv E_0 \cos(\omega t) \equiv -\nabla \Phi$  along $Z$-axis, and director ${\bm n}$ along $X$-axis. The flexoelectric polarization  $\bm P_{fl}$ is defined by Eq. (\ref{weakc7}). This expression can be rewritten in the equivalent form as
\begin{eqnarray}
\bm P_{fl} \, =  \, \zeta\, \bm n (\nabla \bm n) \, + \,  e_3 \, \partial_k (n_k \bm n  ) \,  , \
\label{Pflt}
\end{eqnarray}
where $\zeta= e_1 -  e_3$.

Having in mind that we will solve the equations for the electric field $\bm E$ (potential $\varphi$), we use the thermodynamic potential
formulated in terms of $\bm E$ (see \cite{PM23}). If one considers electric field $\bm E$ as an independent variable, the corresponding thermodynamic potential is
\begin{eqnarray}
U-\frac{1}{4\pi} \bm E \bm D= -\frac{1}{8\pi}\bm E \hat\epsilon \bm E-\bm P_{fl} \bm E
+\frac{1}{2\rho}J^2 +F_F +U_0(S,\rho),
\label{poten1} \\
d\left[U-\frac{1}{4\pi} \bm E \bm D\right]
=-\frac{1}{4\pi} \bm D d \bm E
+Td S+\mu d \rho + \bm v d \bm J
+\frac{\partial U}{\partial \bm n}d\bm n
+\frac{\partial U}{\partial (\partial_i\bm n)}d(\partial_i\bm n) .
\nonumber
\end{eqnarray}

Then (writing all variables in SI units) the components of the electric displacement field $\bm D$ in the linear approximation  can be written as
\begin{eqnarray}
D_{0 x}= -\epsilon_\parallel \epsilon_0 \partial_x \Phi
+(\epsilon_\parallel-\epsilon_\perp) \epsilon_0 n_z E(t) + \zeta (\partial_y n_y+\partial_z n_z),
\nonumber \\
D_{0 y} = -\epsilon_\perp \epsilon_0 \partial_y \Phi, \quad
D_{0 z} = \epsilon_\perp \epsilon_0 E(t)-\epsilon_\perp \epsilon_0 \partial_z \Phi \, , \
\label{lDi}
\end{eqnarray}

in turn, the dynamic equations for the components of $\bm D$ takes the following form
\begin{eqnarray}
-\partial_t [\epsilon_\parallel \epsilon_0 \partial_x^2 \Phi
+  \epsilon_\perp \epsilon_0(\partial_y^2+\partial_z^2)\Phi]
\nonumber \\
+\partial_t \partial_x[
\Delta \epsilon \epsilon_0 n_z E(t)+ \zeta (\partial_y n_y+\partial_z n_z)]
\nonumber \\
=\sigma_\parallel \partial_x^2 \Phi +\sigma_\perp(\partial_y^2+\partial_z^2)\Phi
-(\sigma_\parallel-\sigma_\perp) E(t) \partial_x n_z \, , \
\label{llinear1}
\end{eqnarray}
where $\Delta\epsilon  = \epsilon_\parallel -\epsilon_\perp$. In the next step, to obtain  the linear dynamic equations for components of the  perturbed director $\bm n$ and velocity, we need to find the components of
the variational derivative of the energy over $\bm n$, $\Xi_y$ and $\Xi_z$. In the linear  approximation:
\begin{eqnarray}
\Xi_y=K\nabla^2 n_y -\zeta \partial_y n_z E(t)+\zeta \partial_x\partial_y\Phi \, , \
\label{Xiy}\\
\Xi_z=K\nabla^2 n_z +\zeta \partial_y n_y E(t) +\Delta \epsilon \epsilon_0 E(t)^2 n_z
-\Delta \epsilon \epsilon_0 E(t)\partial_x \Phi +\zeta \partial_x\partial_z\Phi \, . \
\label{Xiz}
\end{eqnarray}

For the incompressible fluid $\nabla {\bm J}=  \rho \nabla {\bm v} = 0$. Then from the equation $\partial_t J_i$ (see \cite{PM23})
we can introduce the variable  $\Pi\,$  for the modified by the change of variables pressure $p$
\begin{eqnarray}
\nabla^2 p -  \zeta E(t) \nabla^2  \partial_x n_z  +   \epsilon_\perp \epsilon_0 E(t) \nabla^2  \partial_z \Phi
+\zeta \nabla^2 \partial_x^2\Phi\,  =\, \nabla^2 \Pi
 \nonumber  \\
 =\, -\zeta E(t) \partial_x^3 n_z +\zeta \partial_x^4\Phi
-K\nabla^2 \partial_x(\partial_y n_y+\partial_z n_z)
-  \epsilon_\perp \epsilon_0 E(t) \partial_z \nabla^2 \Phi.
\label{llinear3}
\end{eqnarray}

Substituting the expressions (\ref{Xiy}-\ref{llinear3}) in the dynamic equations for nematic director and velocity, we find in the linear approximation \cite{PM23}
\begin{eqnarray}
(\rho\partial_t -\eta \nabla^2)v_x = -\partial_x \Pi-\zeta E(t) \partial_x^2 n_z +\zeta \partial_x^3 \Phi,
\label{lvx} \\
(\rho\partial_t- \eta \nabla^2)v_y =-\partial_y \Pi -K\nabla^2 \partial_x n_y ,
\label{lvy} \\
(\rho\partial_t -\eta \nabla^2)v_z=-\partial_z \Pi -K\nabla^2 \partial_x n_z
- \epsilon_\perp E(t) \nabla^2\Phi \ , \
\label{lvz}
\end{eqnarray}
\begin{eqnarray}
\partial_t n_y=\partial_x v_y +\gamma^{-1}\,
\Big[ K\nabla^2 n_y -\zeta \partial_y n_z E(t) +\zeta \partial_x\partial_y\Phi \Big]  \, , \
\label{lny}\\
\partial_t n_z =\partial_x v_z +\gamma^{-1}\,
\Big[ K\nabla^2 n_z +\zeta \partial_y n_y E(t) +\Delta \epsilon E(t)^2 n_z
\nonumber \\
-\Delta \epsilon E(t)\partial_x \Phi +\zeta \partial_x\partial_z\Phi \Big]  \ . \
\label{lnz}
\end{eqnarray}
The equations (\ref{lvx}-\ref{lnz}) constitute a complete set for $\Phi,v_y,v_z,n_y,n_z$ with $\Pi$ defined in Eq. (\ref{llinear3}).
Let note, that  equation for velocity can be rewritten in the form:
\begin{eqnarray}
(\rho\partial_t -\eta \nabla^2) {\bm v}
= - \nabla \Pi + {\bm f} \,
 \ , \
\label{NvSt}
\end{eqnarray}
where
\begin{eqnarray}
f_x \, = \,-\zeta E(t) \partial_x^2 n_z \,
+\zeta \partial_x^3 \Phi \,  \ ,
\nonumber \\
f_y \, = \,-K\nabla^2 \partial_x n_y \,  \ ,
\nonumber \\
f_z \, = \,-K\nabla^2 \partial_x n_z
-  \epsilon_\perp \epsilon_0 E(t) \nabla^2\Phi \,
\, . \
\label{fi0}
\end{eqnarray}
To  exclude the pressure from above equations we apply the $curl$ operation to the velocity equation  Eq. (\ref{NvSt})
\begin{eqnarray}
(\rho\partial_t -\eta \nabla^2) (\partial_y v_z - \partial_z v_y)
=  \partial_y f_z - \partial_z f_y  \, ,
\nonumber \\
(\rho\partial_t- \eta \nabla^2) (\partial_z v_x - \partial_x v_z)
= \partial_z f_x - \partial_x f_z \, ,
\nonumber \\
(\rho\partial_t -\eta \nabla^2)  (\partial_x v_y - \partial_y v_x)= \partial_x f_y - \partial_y f_x \, . \
\label{NvSt1}
\end{eqnarray}

If $\lambda \neq 1$, then one can obtain more complicated equations for $n_y,\ n_z$, which read as \cite{PM23}
\begin{eqnarray}
\partial_t n_y=\partial_x v_y +\frac 1{\gamma}
\left[ K\nabla^2 n_y -\zeta \partial_y n_z \,
 E(t) +\zeta \partial_x\partial_y\Phi\right] - \frac{1 - \lambda}{2}\, \big(\partial_y v_x + \partial_x v_y \big)\, ,
\label{llinear43} \\
\partial_t n_z =\partial_x v_z +\frac{1}{\gamma}
\Big[ K\nabla^2 n_z +\zeta \partial_y n_y E(t)
+\Delta \epsilon \epsilon_0 E(t)^2\, n_z
\nonumber \\
-\Delta \epsilon \epsilon_0 E(t)\partial_x \Phi
+\zeta \partial_x\partial_z\Phi \Big] - \frac{1 - \lambda}2 \, \big(\partial_z v_x + \partial_x v_z \big)\, . \
\label{llinear44}
\end{eqnarray}

With three different Frank moduli we have to replace the nematic elasticity terms in all equations as follows
\begin{eqnarray}
K \nabla^2 n_y \to (K_1 \partial_y^2 +K_2 \partial_z^2 +K_3 \partial_x^2)n_y
+(K_1-K_2) \partial_y \partial_z n_z \ , \
\label{Frankny} \\
K \nabla^2 n_z \to (K_1 \partial_z^2 +K_2 \partial_y^2 +K_3 \partial_x^2)n_z
+(K_1-K_2) \partial_y \partial_z n_y \ . \
\label{Franknz}
\end{eqnarray}

In the case of a nonzero flexoelectric coefficient $e_3$, see Eq. (\ref{Pflt}), some additional terms appear in Eqs. (\ref{linear1}),(\ref{linear2} ), (\ref{linear4}). For example, in Eqs. (\ref{linear2})  the following contributions  should be included into the r.h.s.
\begin{eqnarray}
\partial_t n_y = \partial_x v_y +\frac{1}{\gamma}
\Big[ K\nabla^2 n_y -\zeta \partial_y n_z E_0 \cos(\omega t) +\zeta \partial_x\partial_y\Phi \,+\,
e_3 \,\partial_x \partial_y \Phi \Big]\,
 \, , \
 \label{disszy} \\
\partial_t n_z  = \partial_x v_z +\frac{1}{\gamma}
\Big[ K\nabla^2 n_z +\zeta \partial_y n_y E_0 \cos(\omega t)
+ \Delta\epsilon \epsilon_0  E_0^2 \cos^2(\omega t) n_z
\nonumber \\
-\Delta\epsilon \epsilon_0 E_0 \cos(\omega t)\partial_x \Phi
+ \zeta \partial_x\partial_z\Phi  \,+\,
e_3 \,\partial_x \partial_z \Phi \Big]
 \ , \
\label{dissz1}
\end{eqnarray}
and  to the left part of Eq. (\ref{linear1})
\begin{eqnarray}
-\partial_t [\epsilon_\parallel  \epsilon_0\partial_x^2 \Phi
+  \epsilon_0\epsilon_\perp  \epsilon_0 (\partial_y^2+\partial_z^2)\Phi] + \partial_t \partial_x[
(\epsilon_\parallel-\epsilon_\perp)  \epsilon_0 n_z E_0 \cos(\omega t) \nonumber \\
+ \zeta (\partial_y n_y+\partial_z n_z)  \,+\, e_3 (\partial_y n_y+\partial_z n_z) ]
\nonumber \\
 = \sigma_\parallel \partial_x^2 \Phi
+ \sigma_\perp(\partial_y^2+\partial_z^2)\Phi
- (\sigma_\parallel-\sigma_\perp)
E_0\cos(\omega t) \partial_x n_z
 \ . \
\label{dissz2}
\end{eqnarray}

\section{Film of finite thickness}
\label{app:finite}

Analyzing the uniform film it is convenient to make Fourier transform in terms of the longitudinal coordinates $x$ and $y$. Doing so, one can rewrite Eq. (\ref{llinear1}) in the form
\begin{eqnarray}
\partial_t [  \epsilon_\parallel  \epsilon_0 q_x^2 \Phi +  \epsilon_\perp  \epsilon_0 ( q_y^2 - \partial_z^2)\Phi]
+ q_x \partial_t[ \Delta \epsilon  \epsilon_0 E(t)\, i\,n_z + \zeta ( - q_y n_y + i\, \partial_z n_z)]
\nonumber \\
= - \sigma_\parallel q_x^2 \Phi + \sigma_\perp(-q_y^2 + \partial_z^2)\Phi
- (\sigma_\parallel-\sigma_\perp) E(t) q_x i\,n_z \, , \
\label{ilinear1}
\end{eqnarray}
in turn, Eq. (\ref{llinear3}) takes the form
\begin{eqnarray}
\, \big( - q^2 + \partial_z^2 \big) \Pi
 =\,\zeta E(t) q_x^3 i\,n_z +\zeta q_x^4\Phi
- K \big( - q^2 + \partial_z^2 \big) q_x( - q_y n_y + i\, \partial_z n_z)\,
\nonumber \\
-\,   \epsilon_\perp  \epsilon_0 E(t) \,\partial_z \big( - q^2 + \partial_z^2 \big)\Phi \, , \
\label{iliPi}
\end{eqnarray}
where $q^2=q_x^2+q_y^2$.
Eqs. (\ref{lvx})--(\ref{lvz}) can be rewritten as
\begin{eqnarray}
(\rho\partial_t -\eta \big( - q^2  + \partial_z^2 \big)) i\,v_x
= q_x \Pi + \zeta E(t) q_x^2 i\,n_z
+\zeta q_x^3 \Phi \,,
\label{ilivx} \\
(\rho\partial_t - \eta \big( - q^2  + \partial_z^2 \big)) i\,v_y
= q_y \Pi
 + K \big( - q^2  + \partial_z^2 \big) q_x n_y \, , \
\label{ilivy} \\
(\rho\partial_t - \eta \big( - q^2  + \partial_z^2 \big))v_z = -\partial_z \Pi
- K \big( - q^2  + \partial_z^2 \big) q_x i\,n_z
-  \epsilon_\perp  \epsilon_0 E(t) \big( - q^2  + \partial_z^2 \big)\Phi \, , \
\label{ilinear2}
\end{eqnarray}
and Eqs. (\ref{lny})--(\ref{lnz}) take the form
\begin{eqnarray}
\partial_t n_y= q_x  i\,v_y +\gamma^{-1}
\Big[ K \big( - q^2  + \partial_z^2 \big) n_y -\zeta q_y i\,n_z E(t) - \zeta q_x\, q_y\Phi \Big]  \, , \
\label{iliny} \\
i\,\partial_t n_z = - q_x v_z +\gamma^{-1}
\Big[ K \big( - q^2  + \partial_z^2 \big) i\,n_z - \zeta q_y n_y E(t)
+\Delta \epsilon  \epsilon_0 E(t) i\,n_z
\nonumber \\
+ \Delta \epsilon   \epsilon_0 E(t) q_x \Phi
 - \zeta q_x\partial_z\Phi \Big]  \ . \
\label{ilinear4}
\end{eqnarray}
Let us emphasize, that Eqs. (\ref{ilinear1}), (\ref{ilivx})--(\ref{ilinear2}), (\ref{iliny})--(\ref{ilinear4}) constitute the transformed complete set for $\Phi,v_y,v_z,n_y,n_z$ with $\Pi$ is defined in Eq. (\ref{iliPi}).

After Fourier and (\ref{subst0}) transformations of the velocity equations we obtain
\begin{eqnarray}
 \ (\rho\partial_t -\eta \nabla^2) ( q_y v_z  + i\,\partial_z v_y)
=  q_y f_z + i \partial_z f_y \, ,
\nonumber \\
 \  (\rho\partial_t- \eta \nabla^2) (i\,\partial_z v_x + q_x v_z)
= i\partial_z f_x + q_x f_z \, ,
\nonumber \\
 \  (\rho\partial_t -\eta \nabla^2)  (q_x  i\,v_y -  q_y  i\,v_x) = iq_x f_y - i q_y f_x  \ . \
\label{NvSt2}
\end{eqnarray}
where transformations of $\bm f$ components have the following form
\begin{eqnarray}
f_x \, = \ {- i(\zeta E(t) q_x^2 i\,n_z + \zeta q_x^3 \Phi)} \ ,
\nonumber \\
f_y \, =  \ { - i (K q_x (-q^2 + \partial_z^2)n_y )} \ ,
\nonumber \\
f_z \, =    \ { K q_x (q^2 - \partial_z^2) i\,n_z
+  \epsilon_\perp  \epsilon_0 E(t) (q^2 - \partial_z^2)\Phi}
\, . \
\label{fi}
\end{eqnarray}

Then using the continuity equation we can exclude $v_x$ as
\begin{eqnarray}
 \ { \  i\,v_x = -( q_y  i\,v_y + \partial_z v_z )/q_x}\
 \ . \
\label{CE}
\end{eqnarray}
Thus we arrive to the following system of the equations for the velocity components $v_y$ and $v_z$:
\begin{eqnarray}
 \rho {q^2} \partial_t v_z  -  \rho  \partial_t\partial_z^2 v_z  =  \eta (- {q^4} v_z  + 2 {q^2} \partial_z^2 v_z - \partial_z^4 v_z)
+  q^2 f_z + i {q_y}\partial_z f_y + i q_x\partial_z f_x \, ,
\label{NvSt4vy} \\
i\, {q^2}\rho\partial_t v_y = i\,\eta (-q^2 + \partial_z^2) q^2 v_y  - q_y\, (\rho\partial_t -\eta (-q^2 + \partial_z^2) ) \partial_z v_z \,+ q_x\,(iq_x f_y - i q_y f_x) \ . \
\label{NvSt4vz}
\end{eqnarray}

The third order derivative of $n_y$ over $z$ appears from the term $\partial_z f_y$ in Eq. (\ref{NvSt4vy}). To exclude this term it is possible to express it from Eq. (\ref{ilinear4}) as
\begin{eqnarray}
\gamma ^{-1}\, K \partial_z^3  n_y = \partial_t \partial_z n_y - q_x \, \partial_z  i\,v_y + \gamma ^{-1}\, \partial_z
\Big(  K q^2 n_y +\zeta q_y i\,n_z E(t) + \zeta q_x\, q_y\Phi \Big)
\ , \
\label{ilinear4ny}
\end{eqnarray}
and to substitute the obtained expression in Eq. (\ref{NvSt4vy}).
As a result, we obtain the following equations for ${v}_y$ and ${v}_z$
\begin{align}
& i\,\rho\,q^2\,  \partial_{t} v_y = \,i\,\eta\,\big(- q^2\,  + \, \partial_{z}^2\big)   q^2\,v_y   -  {q}_y \Big(\rho\,\partial_t - \,\eta\,\big(-{q}^2 + \, \partial_{{z}}^2\big) \Big) \, \partial_{z}{v}_z \,
 \nonumber \\
 &- \,  {\zeta}\, {q}_x^3  {q}_y  \,\Big({E}_0 \cos(2 \pi \tau)\, i\,n_z +  \, {q}_x \,{\Phi}\Big) + \,{K_2}\, {q}_x^3 \, \partial_{z}^2 n_y   -  \, {q}_x^3 \Big( K_1 {q}_y^2 + {K_3}\,{q}_x^2\Big) n_y +  \,\Big(K_1 - {K_2}\Big) {q}_x^3 {q}_y i\,\partial_z n_z \ , \qquad
\label{1lvy}\\
& \rho\,  \big( q^2 -   \,\partial_{{z}}^2 \big) \, \partial_{t}{v}_z  =   -\,\eta\,\big(q^2 -   \,\partial_{{z}}^2 \big)^2 {v}_z  + \,q^2\, \Big\{{q}_x\,\Big({K_2}\,{q}_y^2+ {K_3}\,{q}_x^2 - {K_1}\,\partial_{{z}}^2\Big) i\,n_z  -  {q}_x \, \Big({K_1} - {K_2}\Big)\, {q}_y \partial_z n_y
\nonumber \\
&+ \,{\epsilon}_\perp\epsilon_0 E_0 \cos(2 \pi \tau)\, \big( q^2 - \,\partial_{{z}}^2 \big) {\Phi}
\Big\}\,
+ \, {q}_x \,\Big\{ \gamma\,{q}_y  \, \partial_{t}\partial_{{z}}n_y + {\zeta}\,{{E}_0}\cos(2 \pi \tau)\,{q}^2  \, i\,\partial_{{z}} n_z + {\zeta}\, {q}_x {q}^2   \partial_{{z}}{\Phi} -
\gamma\,{q}_x {q}_y\,i\,\partial_{{z}} v_y \Big\}
 \,  .
\label{1lvz}
\end{align}

\section{Comparison}
\label{app:comparison}

Here we present the comparison of the computation results of the flexoelectric instability for  unbounded nematic and for the film of finite thickness (as an illustrative  example we chose $d=7.62~\mu m$). The wave vector $q_z$ for the unbounded nematic was equal to $q_z=\pi/d$. These two cases have been studied with identical material parameters. The following parameters were the same in all calculations: $\omega= 2\pi \cdot 500~s^{-1}$, $\sigma_\perp = 10^{-9}$ $\Omega^{-1} \cdot m^{-1}$ ($10~s^{-1}$), $\Delta \sigma = - 0.2 \sigma_\perp $, $K_1 = 7 $ pN, $K_2 = 5 $ pN, $K_3 = 5 $ pN,  ${\epsilon}_\perp = 14 $, $\Delta {\epsilon} = - 3$, $\gamma = 0.066\, Pa\cdot s$. Parameters $\zeta$ and $\eta$ were different, corresponding values are presented in the headers of the Tables.  The results presented in Tables $I, II, IV$ were obtained at calculation during a period, whereas for Table $III$ calculation was done for $150$ periods of external field.

The tables illustrate the correspondence between the results for the unbounded nematic and for the film of finite thickness for all possible types of the instability. The obtained practically perfect agreement is a result of the smallness of $q_z$ in comparison with the lateral wave vector.

\begin{table}[h]
\caption{Amplification factor of the critical mode $\Lambda(q_{x}, q_{y})$ in the vicinity of the main maximum for the unbounded nematic  and for the film of finite thickness. The static stripes case at $\zeta = 2.79\cdot 10^{-11}$ ${C}\cdot m^{-1}$, $\eta = 0.055\, Pa\cdot s $. }
\label{Comp1}\centering%
\begin{tabular}{clccc}
\toprule%
  $E_{0} \ $ &$\ $ $(q_{x}, q_{y})$ & $\ \ $ $ \Lambda \ $ & $\ \ $ $\Lambda \  $\\
  $\ $ ($V\,\mu m^{-1}$) $\ $ & ($\mu m^{-1}, \mu m^{-1}$) & $\ $ unbounded  $\ $ & $\ $ film\\
    $\ $  $\ $ & $\ $ & nematic &\\ \hline
   & (0, 1.56)  & 1.00031  & 1.00022 \\
  & (0, 1.18)  & 0.995  & 0.995 \\
   & (0, 1.8)  & 0.998  & 0.998\\
  $\,$ 4$\ $& (0.052, 1.84) &  0.9965 &  0.9964\\
    & (0.052, 1.39)   & 0.9987  & 0.9987\\
    & (0.21, 1.53)  & 0.9905  & 0.9906 \\
     & (0.21, 1.84)  &  0.989  &  0.989 \\
   \toprule%
   \end{tabular}
\end{table}

\begin{table}[h]%
\caption{Amplification factor of the critical mode $\Lambda(q_{x}, q_{y})$ in the vicinity of the main maximum for unbounded nematic and for the film of finite thickness. The case of static periodic two-dimensional director pattern at $\zeta \approx 6.23\cdot 10^{-11}$ ${C}\cdot m^{-1}$, $\eta = 0.055\, Pa\cdot s $.}
\label{Comp2}\centering%
\begin{tabular}{clccc}
\toprule%
  $E_{0} \ $ &$\ $ $(q_{x}, q_{y})$ & $\ \ $ $ \Lambda \ $ & $\ \ $ $\Lambda \  $\\
  $\ $ ($ V \mu m^{-1}$) $\ $ & ($\mu m^{-1}, \mu m^{-1}$) & $\ $ unbounded  $\ $ & $\ $ film\\
    $\ $  $\ $ & $\ $ & nematic &\\ \hline
  & (1.53, 5.25)   & 1.0002 & 0.9993 \\
 & (1.51, 5.25)   & 0.9998  & 0.9986 \\
  & (1.52, 5.25)  & 1.00009 & 0.9991 \\
 $\ $ $\,$ 4$\ $ & (1.52, 5.27)  & 1.00014 & 0.999 \\
    & (1.53, 5.32) &  0.9995 & 0.9982 \\
    & (1.544, 5.20)  & 0.9999 & 0.9988\\
     & (1.544, 5.31)  & 0.9996  & 0.99865 \\
     & (1.56, 5.22)  & 1.0001 & 0.99915  \\
     & (1.56, 5.34)  & 0.9982 & 0.9971 \\
     & (1.76, 5.86)  & 0.99  & 0.9907  \\
           \toprule%
   \end{tabular}
\end{table}

\begin{table}[h]%
\caption{Amplification factor of the critical mode $\Lambda(q_{x}, q_{y})$ in the vicinity of the main maximum for unbounded nematic and for the film of finite thickness. The case of static periodic two-dimensional director pattern for the solution for the film at $\zeta \approx 6.23\cdot 10^{-11}$ ${C}\cdot m^{-1}$, $\eta = 0.055\, Pa\cdot s $. }
\label{Comp22}\centering%
\begin{tabular}{clccc}
\toprule%
 $E_{0} \ $ &$\ $ $(q_{x}, q_{y})$ & $\ \ $ $ \Lambda \ $ & $\ \ $ $\Lambda \  $\\
  $\ $ ($ V \mu m^{-1}$ ) $\ $ & ($\mu m^{-1}, \mu m^{-1}$) & $\ $ unbounded  $\ $ & $\ $ film\\
    $\ $  $\ $ & $\ $ & nematic &\\ \hline
   & (1.53, 5.25)   & 1.0002 & 1.0001\\
  & (1.51, 5.25)   & 0.9998 & 0.9996 \\
 & (1.52, 5.25)  & 1.00009 & 0.9999 \\
  $\ $ $\,$ 4$\ $ & (1.52, 5.27)  & 1.00014 & 0.99997\\
   & (1.53, 5.32) &  0.9995 & 0.9994 \\
     & (1.544, 5.20)  & 0.9999  & 0.9998 \\
     & (1.544, 5.31)  & 0.9996  & 0.99955\\
    & (1.56, 5.22)  & 1.0001 & 1.00008 \\
    & (1.56, 5.34)  & 0.9982   & 0.998\\
    & (1.76, 5.86)  & 0.99 & 0.99\\
                      \toprule%
\end{tabular}
\end{table}

\begin{table}[tbp]%
\caption{Amplification factor of the critical mode $\Lambda(q_{x}, q_{y})$ in the vicinity of the main maximum for unbounded nematic and for the film of finite thickness. The case of oscillatory two-dimensional director pattern at $\zeta = 2.79\cdot 10^{-11}$ ${C}\cdot m^{-1}$, $\eta = 0.023\, Pa\cdot s $.}
\label{Comp3}\centering%
\begin{tabular}{clcccc}
\toprule%
 $E_{0} \ $ &$\ $ $(q_{x}, q_{y})$ & $\ \ $ $ \Lambda \ $ & $\ \ $ $\vert \Lambda \vert $& $\ \ $ $\Lambda \  $ & $\ \ $ $\vert \Lambda \vert $\\
  $\ $ ($ V \mu m^{-1}$ ) $\ $ & ($\mu m^{-1}, \mu m^{-1}$) & $\ $ unbounded  $\ $ & $\ $ unbounded  $\ $ & $\ $ film & $\ $ film\\
    $\ $  $\ $ & $\ $ & nematic & nematic &  &\\ \hline
 & (0.99, 1.87)  & 0.997  $\pm$ i\, 0.075  &$\ $ 1.00012 & 1.00085  $\pm$ i \,0.0789509 & $\ $ 1.00396\\
      & (1.155, 1.98) &  0.993  $\pm$ i\,  0.088 &   0.997 & 0.997  $\pm$ i \,0.092  & 1.0014\\
     & (0.924, 1.81)  & 0.997 $\pm$ i\, 0.07 & 0.9997 & 1.0007  $\pm$ i \,0.071   &  1.0032 \\
   $\ $ 3.75   & (0.924, 1.923)  & 0.996  $\pm$ i\, 0.071\,  & 0.998  & 0.999  $\pm$ i \, 0.075  &  1.00195 \\
     & (0.924, 2.036)  & 0.992  $\pm$ i\, 0.072 & 0.995 & 0.996  $\pm$  i \, 0.0726  & 0.9988\\
    & (0.957, 1.753)  & 0.997  $\pm$ i\,  0.0716  & 0.9999  & 1.0007  $\pm$ i\, 0.0746 &  1.0035 \\
    & (0.957, 1.81)  & 0.9975  $\pm$ i\, 0.0721  & 1.0001 & 1.0012  $\pm$  i\, 0.073  & 1.0039  \\
    & (0.924, 1.75)  & 0.9974  $\pm$ i\, 0.0693& 0.9998 & 1.0012  $\pm$  i\, 0.073  & 1.0039 \\
     & (1.12, 1.92)  &  0.994  $\pm$ i\, 0.085 &  0.998& 0.998  $\pm$ i\, 0.087 &  1.002 \\
     & (1.12, 2.205)  &  0.99 $\pm$ i\, 0.088  & 0.994 & 0.994 $\pm$ i\, 0.092  & 0.998\\
     & (0, 0.34) &  0.982  &  0.982 &  0.982  &  0.982 \\
& (0, 1.923)  & 0.982  &  0.982 &  0.982  &  0.982\\
 & (1.55, 0.396)  & 0.981 & 0.981  & 0.985 & 0.985 \\
   \end{tabular}
\end{table}

\end{document}